\documentclass[%
prx,amsmath,amssymb,twocolumn,
 superscriptaddress,
 longbibliography,
 aps
]{revtex4-2}
\usepackage{xcolor}
\usepackage{amsmath,amssymb,bbold,bm}
\usepackage{graphicx}
\usepackage{mathdots}
\usepackage{cancel}
\usepackage{comment}
\usepackage{currfile}

\newcommand{\sbraket}[1]{\langle #1 \rangle}

\newcommand{\be}{\begin{equation}}
\newcommand{\ben}{\begin{equation*}}
\newcommand{\ee}{\end{equation}}
\newcommand{\een}{\end{equation*}}
\newcommand{\bs}{\begin{split}}
\newcommand{\es}{\end{split}}
\newcommand{\bmx}{\begin{array}}
\newcommand{\emx}{\end{array}}
\newcommand{\bea}{\begin{eqnarray}}
\newcommand{\bean}{\begin{eqnarray*}}
\newcommand{\eea}{\end{eqnarray}}
\newcommand{\eean}{\end{eqnarray*}}
\newcommand{\dg}{^{\dagger}}
\newcommand{\dn}{^{\vphantom{\dagger}}}

\newcommand{\ua}{\uparrow}
\newcommand{\da}{\downarrow}
\newcommand{\Ua}{\Uparrow}
\newcommand{\Da}{\Downarrow}
\newcommand{\bb}[1]{\mathbb{#1}}

\newcommand{\andd}{\qquad\text{and}\qquad}

\newcommand{\eps}{\epsilon}

\newcommand{\pref}[1]{(\ref{#1})}

\newcommand{\intob}[1]{\int_{0}^{\beta}{#1}}

\newcommand{\abs}[1]{\left\vert #1 \right\vert}
\newcommand{\bra}[1]{\left\langle #1 \right\vert}
\newcommand{\ket}[1]{\left\vert #1\right\rangle}

\newcommand{\braket}[1]{\left\langle #1\right\rangle}

\newcommand{\mat}[1]{\left(\bmx{cc}#1\emx\right)}

\usepackage{enumitem}
\usepackage{color}
\usepackage{graphicx}
\usepackage{bm}
\usepackage{hyperref}
\usepackage{braket}

\definecolor{eggplant}{RGB}{148,33,147}
\definecolor{royalblue}{RGB}{4,51,255}
\definecolor{orange}{RGB}{255,147,0}
\definecolor{fern}{RGB}{79,143,0}

\newcommand{\eq}[1]{Eq.~\eqref{#1}}
\newcommand{\pmat}[1]{\begin{pmatrix}#1\end{pmatrix}}

\begin{document}

\title{Matrix Product Study of Spin Fractionalization in the 1D Kondo Insulator}

\author{Jing Chen}
\email{jchen@flatironinstitute.org}
\affiliation{Center for Computational Quantum Physics, Flatiron Institute, 162 5th Avenue, New York, NY 10010, USA}

\author{E.\ Miles Stoudenmire}
\email{mstoudenmire@flatironinstitute.org}
\affiliation{Center for Computational Quantum Physics, Flatiron Institute, 162 5th Avenue, New York, NY 10010, USA}

\author{Yashar Komijani}
\email{komijani@uc.edu}
\affiliation{Department of Physics, University of Cincinnati, Cincinnati, Ohio 45221-0011, USA}

\author{Piers Coleman}
\email{coleman@physics.rutgers.edu}
\affiliation{Center for Materials Theory, Rutgers University,
Piscataway, New Jersey, 08854, USA}
\affiliation{Hubbard Theory Consortium, Department of Physics, Royal Holloway, University of London, Egham, Surrey TW20 0EX, UK}

\date{\today}

\begin{abstract}
The Kondo lattice is one of the classic examples of strongly correlated electronic systems. We conduct a controlled study of the Kondo lattice in one dimension, highlighting the role of excitations
created by the composite fermion operator. Using time-dependent matrix-product-state methods we compute various correlation functions and contrast them with both large-N mean-field theory and the strong-coupling expansion. We show that the composite fermion operator creates long-lived, charge-e and spin-1/2 excitations, which cover the low-lying single-particle excitation spectrum of the system. Furthermore, spin excitations can be thought to be composed of such fractionalized quasi-particles with a residual interaction which tend to disappear at weak Kondo coupling.
\end{abstract}

\maketitle

\section{\label{intro}Introduction } Kondo insulators are an important
class of quantum material, which historically, foreshadowed the
discovery of heavy fermion metals and superconductors \cite{Menth_PRL1969}.  These
materials contain localized d or f-electrons, forming a lattice of
local moments, immersed in the sea of conduction electrons \cite{Mott74,Doniach77,Fisk92,Riseborough2010}.  Remarkably, even though the high temperature
physics is that of a metallic half-filled band, at low temperatures,
these materials transition from local moment metals, to paramagnetic
insulators.  In the 1970s, theorists came to appreciate that the
origin of this behavior derives from the formation of local singlets
through the action of an antiferromagnetic exchange interaction
between electrons and magnetic
moments \cite{Mott74,Doniach77,kasuya1955,lacroix79}, a model
known as the Kondo
lattice Hamiltonian. 

The Kondo lattice model 
\begin{eqnarray} 
H = -t_c\sum_{ \langle i, j \rangle \sigma } ( c^\dagger_{i,\sigma}
c_{j,\sigma} + {\rm H.c} )+ 
J \sum_j (c^\dagger_{j}\, \vec{\sigma}\, c_{j}) \cdot \vec{S}_{j}
\label{eqn:HK}
\end{eqnarray}
contains a tight-binding model of mobile electrons coupled
antiferromagnetically to a lattice of local moments via a Kondo
coupling constant $J$. The deceptive simplicity of this model hides
many challenges. Perturbative expansion in $J$, reveals that the Kondo
coupling is marginally relevant, scaling to strong-coupling at an
energy scale of the order of Kondo temperature \mbox{$T_K\sim W e^{-1/J\rho}$}. 
Moreover, the localized moments, with a two-dimensional
Hilbert space, do not allow a traditional Wick expansion of the
Hamiltonian, impeding the application of a conventional
field-theoretic methods. The strong-coupling limit of this
Hamiltonian, in which $J$ is much larger than the band-width, $J/t_c\gg
1$ provides a useful caricature of the Kondo insulator as an
insulating lattice of local singlets. 
In the 1980s
\cite{Coleman84,Read84,Auerbach86,Millis87}, new insight into the
Kondo lattice was obtained from the large-$N$
expansion. Here, extending the spin symmetry from the SU(2)
group, with two-fold spin degeneracy, to a family of models with $N$
fold spin-degeneracy allows for an expansion around the large-$N$
limit in powers of $1/N$.  The physical picture which emerges from the
large-$N$ expansion accounts for the insulating behavior in terms of a
\emph{fractionalization} of the local moments into spin-1/2
excitations, $\vec S_j\rightarrow f\dg_{j\alpha} (\vec
\sigma/2)_{\alpha \beta}f_{j\beta}$ which hybridize with conduction
electrons \cite{lacroix79,Read84,Auerbach86,Millis87,Coleman87b} to
form a narrow gap insulator.  However, the use of the large-$N$ limit
provides no guarantee that the main conclusions apply to the
most physically interesting case of $N=2$.

In this paper we use matrix product state methods to examine the physics of
the one dimensional Kondo insulator.  Our work is motivated by a
desire to explore and contrast the predictions of the strong coupling and large-$N$
descriptions with a computational experiment, taking into account the following
considerations:
\begin{itemize}
    \item  Traditionally, Kondo insulators are regarded as an adiabatic evolution of a band-insulating ground-state of a half-filled Anderson lattice model.  We seek to understand the insulating behavior, which is akin to a ``large Fermi surface'',  from a purely Kondo lattice perspective, without any assumptions as to the electronic origin of the local moments. 
     \item  What are the  important differences between the excitations of a half-filled Kondo insulator and a conventional band insulator?  
     \item Many aspects of the Kondo lattice suggested by the large-$N$ expansion, most notably the formation of composite fermions and the associated fractionalization of the spins, have not been extensively examined in computational work.      
     In this respect, our work complements the recent study of Ref.\,\cite{Danu2021}, highlighting the mutual independence of the conduction electron and composite fermions through the matrix structure of the electronic Green's function. We extend this picture even further by examining the dynamical spin susceptibility, providing evidence for fractionalization of the spin into a continuum of quasi-particle excitations.

\end{itemize}

\subsection{Past studies}\label{}

Our work builds on an extensive body of earlier studies of the 1D Kondo
lattice that we now briefly review. 
The  ground-state phase diagram of this model was first established by Tsunetsugu, Sigrist, and Ueda \cite{Tsunetsugu}, who established the stability of the insulating phase for all ratios of $J/t_c$, while also demonstrating that upon doping, the 1D Kondo insulator becomes a ferromagnet. 
More recently, the 1D KL has been studied using Monte Carlo \cite{Fye90,Troyer93,Raczkowski19}, density matrix renormalization group (DMRG) \cite{Yu93,Sikkema97,McCulloch99,Shibata99,Peters12}, bosonization \cite{Zachar96,Tsvelik19}, strong-coupling expansion \cite{Sigrist97} and exact diagonalization \cite{Basylko08}. 
Additionally, renormalization and Monte-Carlo methods have also been used to examine the p-wave version of the 1D Kondo lattice, which exhibits topological end-states \cite{Lobos15,Zhong2017,Zhong2018}.

The Kondo insulator can be driven metallic by doping,
which leads to a closing of charge and spin gaps, forming a Luttinger
liquid with parameters that evolve with doping and $J/t_c$
\cite{Shibata99,Basylko08}. Both the insulating phase at half-filling
and the doped metallic regime are non-trivial, as the $k_F$, extracted
from spin and charge correlation functions corresponds to a large
Fermi surface, which counts both the electrons and spins $v_{\rm FS}/\pi=n_e+1$. The weak-coupling $J/t_c \ll 1$
regime at finite doping continues to be paramagnetic, but the
strongly-coupled $J/t_c\gg 1$ regime becomes a metallic ferromagnetic state for
infinitesimal doping. In this regime, the spin-velocity goes to zero,
characteristic of a ferromagnetic state \cite{Khait5140}, as inferred
from spin susceptibility.

The excitation spectrum of a one dimensional Kondo lattice at half-filling
was first studied by Trebst et al.\,\cite{Trebst06} who employed a strong coupling
expansion in $J/t_c$ to examine the one and two-particle spectrum. 
Their studies found that beyond $t_c/J>0.4$, 
the minimum in the quasi-hole spectrum 
shifts from $k=\pi$ to $k<\pi$. Furthermore, they
extracted the quasi-particle weights showing that $Z\to 0$ right at
$t_c/J=0.4$ when the dispersion is flat around $k\sim\pi$. Smerat et
al.\,\cite{Smerat09} used DMRG to compute the quasi-particle energy
and lifetime to verify these results and extend them to partial
filling. They pointed out that the exchange of spin between conduction
electrons and localized moments leads to formation of
``spin-polarons'', here referred to as ``composite fermions''.


\subsection{Motivation and summary of results }\label{}

The appearance of an insulator in
a half-filled band goes beyond
conventional band-theory and requires a new conceptual framework. 
A large body of work, dating back to the 1960s recognized that
there are two ways to add an electron into a system containing localized moments \cite{Appelbaum1966,Anderson1966,Pustilnik01,Maltseva2009}, either by direct ``tunneling'' an
electron into the system, formally by acting on the state with the
conduction electron creation operator $c\dg_\sigma$, or 
by ``cotunneling'' via 
the simultaneous addition of an electron and a flip of the local moment at the same site $F\dg_\sigma\sim c\dg_{\bar \sigma }S_{\sigma \bar \sigma }=
c\dg_{\bar\sigma }\ket{\sigma}\bra{\bar\sigma}
$.  Both processes change the charge 
by $e$ and the spin by one half.  The object created by $F\dg $ has also alternately referred to as  a ``composite fermion''  or a ``spin-polaron'' \cite{Smerat09}. Here we will employ the former terminology, introducing 
\be
F\dg_{\beta}=\frac{2}{3}\sum_{\alpha=\ua,\da}c\dg_{\alpha}\vec\sigma_{\alpha\beta}\cdot \vec{S} \label{eqn:FS}.
\ee
as the composite fermion creation operator: $F\dg_{\beta}$ transforms as a charge $e$ and spin $S=1/2$ fermion, and with the above normalization the expectation value of its commutator with the conduction electron operator vanishes $\langle\{c_{\alpha},F\dg_{\beta}\}\rangle =0$, while that of commutator with itself is unity $\langle\{F_{\alpha},F\dg_\beta\}\rangle =\delta_{\alpha\beta}$, in the strong Kondo coupling limit.

Co-tunneling lies at the heart of the Kondo problem, and insight into its physics can be obtained by observing that in the
interaction, the object that couples to electron in the
Kondo interaction is a \emph{composite fermion}, for 
\begin{equation}\label{}
J (c^\dagger_{j}\, \vec{\sigma}\, c_{j}) \cdot \vec{S}_{j}  \equiv \frac{3J}{4}
[{F\dg_{j\sigma } } c_{j\sigma }+ c\dg_{j\sigma}F_{j\sigma }].\label{eq2}
\end{equation}
In certain limits, such as the large
$J$ limit and  large-$N$ limit, $F$ behaves as a physically independent
operator, suggesting that the Kondo effect involves a hybridization of the conduction electrons with an emergent, fermionic field.  The large-$N$ limit accounts for the emergence of the
independent composite fermions as a consequence of a fractionalization of
the local moments, and in this limit, both the composite fermion 
and the local moments are described in terms of a single $f$-electron
field, 
\begin{eqnarray}\label{}
F_{j}&\sim&f_{j},\cr{\vec{S}_{j}}&\rightarrow &f\dg_{j\alpha }\left(\frac{\vec{\sigma }}{2} \right)_{\alpha \beta }f_{j\beta}.
\end{eqnarray}

Though the Kondo Lattice is a descendent of the Anderson lattice, it
exists in its own right. In particular, rather than the
four-dimensional Hilbert space of an electron at each site, the spins have
a two-dimensional Hilbert space.  If there are ``f" electrons they are
by definition, $Z=0$ quasiparticles, as there is absolutely no
localized electron Hilbert space.  Field theory and DMRG of single
impurity provide a clue: the presence of 
many body poles in the conduction self-energy
can be interpreted in a dual picture as the hybridization of the
conduction electrons with fractionalized spins.

One of the key objectives
of this work is to 
shed light on the quantum mechanical interplay between the composite
fermion, the conduction electron and the possible fractionalization of
local moments in a spin-1/2 1D Kondo lattice (1DKL). This is achieved by carrying out a new set of calculations of the dynamical properties of the Kondo lattice while also comparing the results with those of large $N$ mean-field theory and strong coupling expansions about the large $J$ limit. In each of these methods, we evaluate the joint matrix Green's function describing the time evolution of the conduction and composite fermion fields following a tunneling or cotunneling event. 

Matrix-product states are ideally suited to one dimensional quantum
problems, permitting an economic description 
of the many-body ground-state with sufficient precision to explore the
correlation functions in the frequency and time domain.   Here, we take
advantage of this method to compute Green's function matrix between conduction electrons and composite fermions and to compare the
spin correlation functions of the local moments and composite
fermions. For simplicity, we limit ourselves to zero temperature $T=0$.
At the one-particle level we find that by analyzing the  Green's
function matrix between $c$ and $F$, we are able to show that these operators
define a hybridized two-band model, in agreement with the large-$N$
limit. The evolution of our computed one-particle spectrum with $t_c/J$ is
consistent with earlier strong-coupling expansions. Rermarkably, the
shift in the minimum of the quasiparticle dispersion seen in the
strong-coupling expansion at $t_c/J=0.4$ \cite{Trebst06} can be qualitatively accounted for in terms
of the evolution of the hybridization between conduction electrons 
and composite fermions. 

Moreover, by calculating the dynamical spin susceptibility
using MPS methods, and comparing the results with mean-field theory, we are able to identify a continuum in the spin excitation spectrum that is consistent with the fractionalization of the local moments into pairs of $S=1/2$ excitations.  Our strong coupling expansion coincides with the matrix product state calculation in the large $J$ limit and we also see signs of the formation of $S=1$ paramagnon bound-states below the continuum: a sign of quiescent magnetic fluctuations.

\section{Model and Methods}

The model we consider is deceptively simple. It is given by the one-dimensional Kondo lattice Hamiltonian
\begin{align} 
H = -t_c\sum_{ i\sigma } ( c^\dagger_{i,\sigma} c_{i+1,\sigma} + \text{h.c.} ) + J \sum_i (c^\dagger_{i}\, \vec{\sigma}\, c_{i}) \cdot \vec{S}_{i} 
\label{eqn:H}
\end{align}
where $c\dg_{i,\sigma}$ creates an electron of spin $\sigma$ at site $i$ and $t$ controls the electron tunneling
between sites. 
The operator $\vec{S}_{i}$ is an immobile $S=1/2$ spin located at site $i$ and 
$(c^\dagger_{i}\, \vec{\sigma}\, c_{i}) \cdot \vec{S}_{i}$
 is a Heisenberg coupling between the spin moment of an 
electron at site $i$ and the spin $\vec{S}_{i}$.
In the limit of large $J/t_{c}$ the half-filled ground-state is composed of a product of Kondo singlets at every site, a state that is
self-evidently an insulator.   The challenge then, is to
understand how this state evolves at finite $J/t_{c}$.

\subsection{Matrix Product State Methods}

\begin{figure}
    \includegraphics[width=0.9\linewidth]{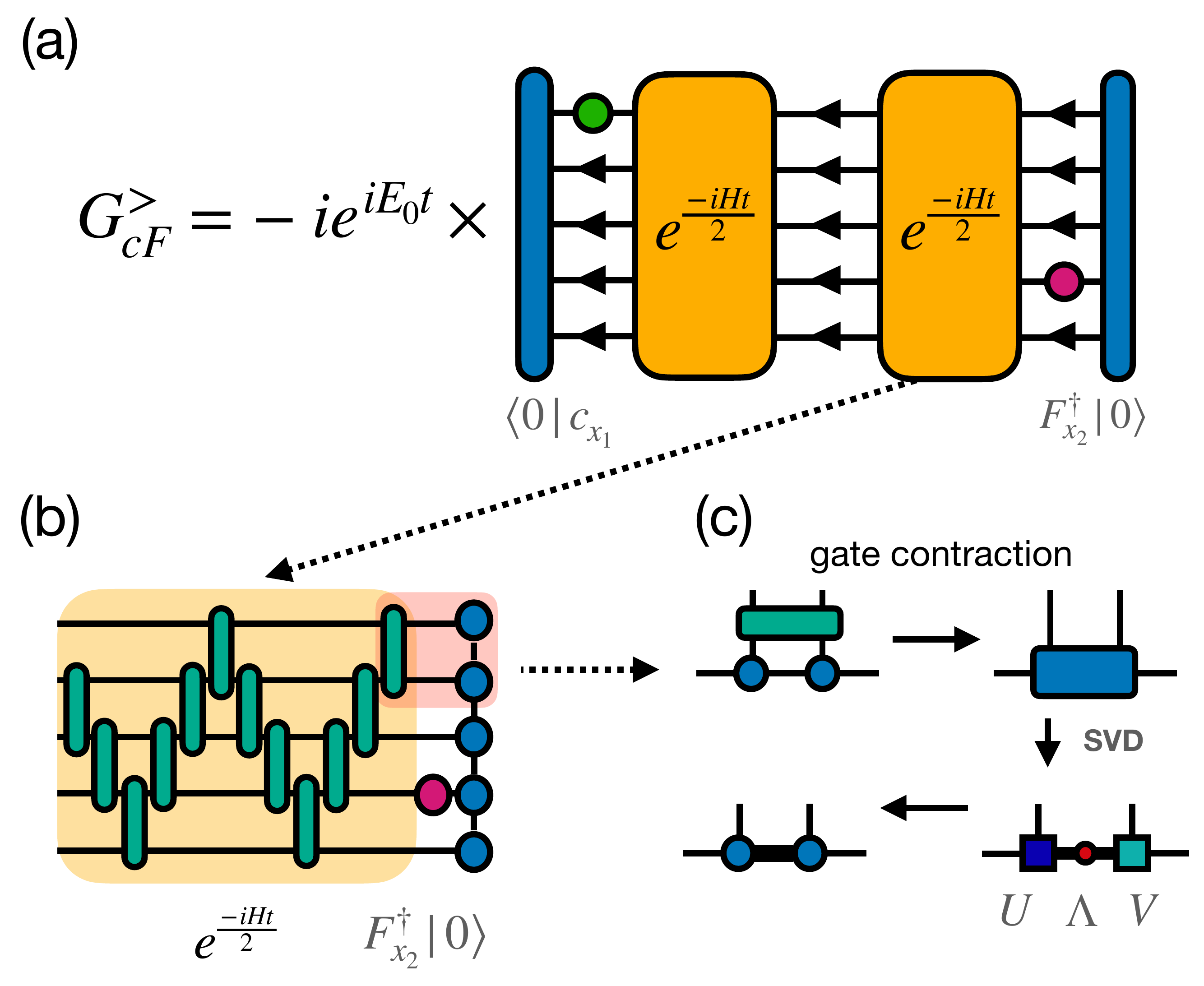}
    \caption{\small Diagrammatic representation of calculating the Green function by MPS methods. The MPS $\ket{0} $ (blue) is the ground state found using  DMRG. The small circles in (a) represent the single-site operators $\ket{F^{\dagger}_{x_2}}$ (magenta circle) and $c_{x_1}$ (green circle). They can be placed at any sites $x_1$ and $x_2$ (though requiring separate computations), giving access to the Green function in real space. The time evolution operator (orange rectangle) is split into two halves, each half approximated by a sequence of unitary gates (dark green rectangles) using a Trotter approximation. The Green functions is found by computing the overlap of the two independent time-evolved wavefunctions. 
         In the bottom right region, we demonstrate the procedures during every step in the time evolvolution (red shaded region). The gate tensors are contracted with MPS tensors, followed by a singular value decomposition (SVD) to reorganize the tensors back into MPS form but with a increased bond dimension.
    }
    \label{fig:tn_diagram}
\end{figure}

The primary tool we will use to study the properties of the Kondo lattice model will be
matrix product state (MPS) tensor networks. An MPS is a highly compressed representation of a 
large quantum state as a contraction of many smaller tensors and is the seminal example of a tensor network.
In contrast to  other numerical or analytical approaches, MPS methods work well for both weakly and
strongly correlated electronic systems and do not suffer from a sign problem as in the case of quantum Monte Carlo
methods. The main limitation of MPS is that they are only efficient for studying one-dimensional or quasi-one-dimensional
systems.

The two key MPS techniques we use in this work are the density matrix renormalization group (DMRG) algorithm for computing ground
states in MPS form \cite{Schollwoeck11}, and the time-evolving block decimation (TEBD) or Trotter gate method for evolving an MPS wavefunction forward in time \cite{Vidal04,Daley04,White04}. Our implementation is based on the ITensor software \cite{itensor}.

Our computational approach is illustrated at a high level in Fig.~\ref{fig:tn_diagram}, using the example of computing 
\mbox{$G^>_{cF} = i\, e^{i E_0 t} \  \langle 0| c_{x_1} e^{-i H t} F^\dagger_{x_2} |0\rangle$}. After computing an MPS representation of the ground state $|0\rangle$ 
using DMRG, we act with $F^\dagger_{x_2}$ on one copy of $\ket{0}$ and with $c^\dagger_{x_1}$ on another copy of $\ket{0}$. The first
copy is evolved forward in time by acting $e^{-i H t/2}$ using a Trotter decomposition of the time evolution operator, and the second is evolved
similarly but acting with $e^{i H t/2}$. Finally, the Green function is computed from the overlap of the resulting MPS. We give additional technical details of our computational approach in Appendix~\ref{MPSGF}.


\begin{figure}
    \centering
    \includegraphics[width=0.7\linewidth]{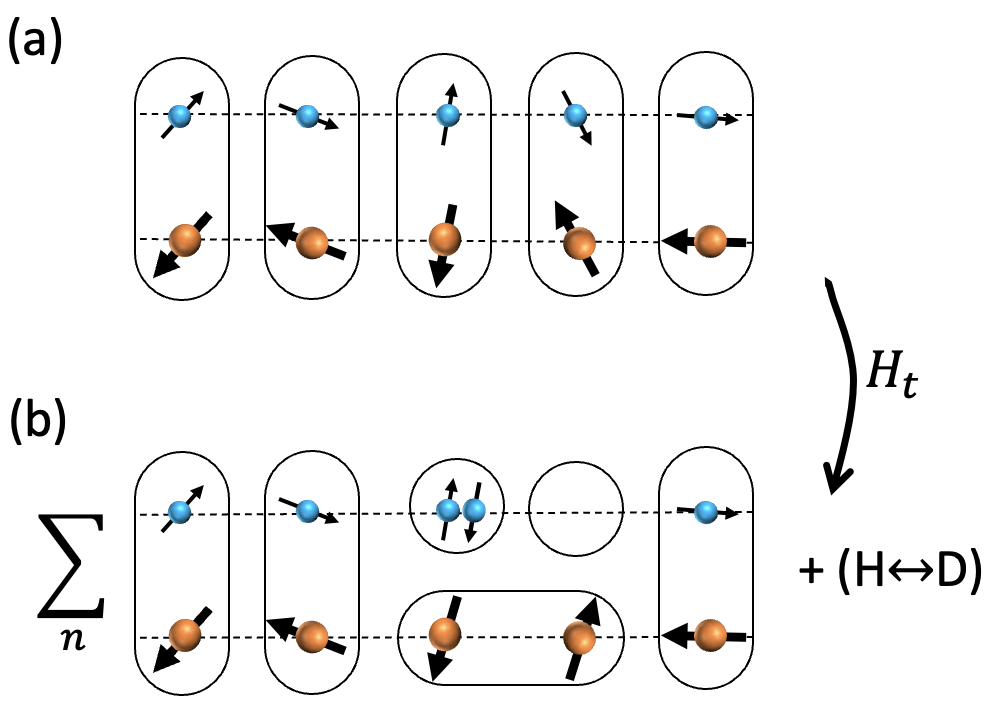}
    \caption{\small Strong coupling diagram. (a) Ground state is comprised of local singlets between spins and conduction electrons. (b) The hopping term in the Hamiltonian creates doublon-holon pairs  whose corresponding spins are together in a singlet states, i.e. charge-0, spin-singlet admixture.}
    \label{fig:fig1}
\end{figure}
\subsection{Strong coupling expansion}
An insight into the nature of ground state and its elementary excitations can be obtained in the strong Kondo coupling limit. At $t_c/J_K=0$ the ground state is a product state of local Kondo singlets
The ground state is
\be
\ket{\phi}_0=\prod_j\ket{K_j}, \qquad \ket{K_j}=\frac{1}{\sqrt 2}\ket{\Ua_{j}\da_j-\Da_{j}\ua_j}
\ee
Here, $\Uparrow$ and $\Downarrow$ refer to the spins (magnetic moments) and $\ua$ and $\da$ refer to conduction electrons. The spin-1/2 excitations corresponding to addition/removal of electrons and spin-1 excitations of changing local singlets into triplets. At finite $t_c/J_K$ electrons hop to nearby sites, creating holon-doublon virtual pairs [Fig.\,\ref{fig:fig1}(b)]. Consequently, the vacuum contains short-lived holon-doublon pairs, which lead to short-range correlations.

\section{Composite Fermions: Single-Particle Properties}
\subsection{More details on the composite fermion operator}
 
To characterize the single-particle excitations of the Kondo lattice, observe that acting on the ground
state by the operators $c\dg_\ua$ and $c^\dagger_\da S^+$ each create charge-1, spin-$1/2$
excitations. However, instead of $c^\dagger_\da S^+$, we will find that the \emph{composite fermion} operator
\be
F\dg_{\beta}=\frac{2}{3}\sum_{\alpha=\ua,\da}c\dg_{\alpha}\vec\sigma_{\alpha\beta}\cdot \vec{S} \label{eqn:F}
\ee
is the more natural operator to consider. One motivation is that $F\dg_{\sigma}$ transforms under the $S=1/2$ 
representation of $SU(2)$. A more intuitive motivation is that the spin-electron interaction term in the Kondo lattice Hamiltonian \eq{eqn:H} can be written as $(\vec{S} \cdot c^\dagger\vec\sigma c) \ \propto\ (F\dg_\sigma c\dn_\sigma+\text{h.c.})$, thus $F\dg_\sigma$ is the operator which couples to electronic excitations. The factor of $2/3$ in \eq{eqn:F} has been chosen so that the commutator (see Appendix \ref{secFcom} for the proof)
\bea
\{F\dn_\alpha,F\dg_\beta\}&=&\delta_{\alpha\beta}\label{FFcomm}\\
&-&\frac{4}{9}\Big[\Big(\vec S\cdot c\dg\vec\sigma c+\frac{3}{2}\Big)\delta_{\alpha\beta}+(\hat n-1)\vec S\cdot\vec\sigma_{\alpha\beta}\Big]\nonumber.
\eea
is unity in the strong coupling limit $J/t_c \gg 1$.
The second line spoils the canonical anti-commutation of $F$ operators, however, in the strong coupling limit $J/t_c \gg 1$ the expectation value of the second term is zero in the ground state, indicating that $\braket{\{F\dn_\alpha,F\dg_{\beta}\}}=\delta_{\alpha\beta}$ has canonical anti-commutation on average. Within the triplet sector, the expectation value of the anticommutator becomes $\delta_{\alpha\beta}/9$ and within the holon/doublon manifold, the expectation value of the anticommutator depends on the state of the magnetic moment. The overlap between the original $c$ and the composite $F$ electrons is
\be
\{c_\alpha,F\dg_\beta\}=\frac{2}{3}\vec S\cdot\vec\sigma_{\alpha\beta}.\label{cFcomm}
\ee
The right-hand side has zero average (but finite fluctuations) in the strong-coupling ground state, suggesting that 
$c$ and $F$ create independent excitations in average. However, $F_\sigma$ and $c_\sigma$ overlap due to quantum fluctuations, motivating us to compute the full Green's function matrix involving both operators to study their associated excitations in a controlled way.

This approximately particle-like behavior of the composite-fermion $F_\sigma$ has strong resemblance to the two-band model of heavy-fermions obtained in the large-$N$ mean-field theory. In such a model the spin is represented using Abrikosov fermions $\vec S=\frac{1}{2}f\dg{\vec\sigma}f$ and the constraint $f\dg f=1$ is applied on average using a Lagrange multiplier. Within mean-field theory, the Kondo interaction leads to a dynamic hybridization between $f$-electrons and $c$-electrons [c.f. Eq.\,\pref{eq2}]
\be
\frac{3}{4}JF\dg_\sigma c\dn_\sigma= Vf\dg_\sigma c\dn_\sigma\label{eq10}
\ee
The similarity of the two results suggests 
$F_\sigma\sim f_\sigma$, 
implying the fact that the spin is fractionalized into spinons. In fact we can define 
\begin{eqnarray}
\vec{S}_F=\frac{1}{2}F\dg\vec\sigma F\label{eqS_F}
\end{eqnarray}

In the rest of this section we will confirm the picture outlined above by computing the full Green function using
two approaches. We first use time-dependent matrix product state techniques on finite systems, then carry out
a strong-coupling analysis to shed further light on the results. 


\begin{figure}
    \includegraphics[width=1\linewidth]{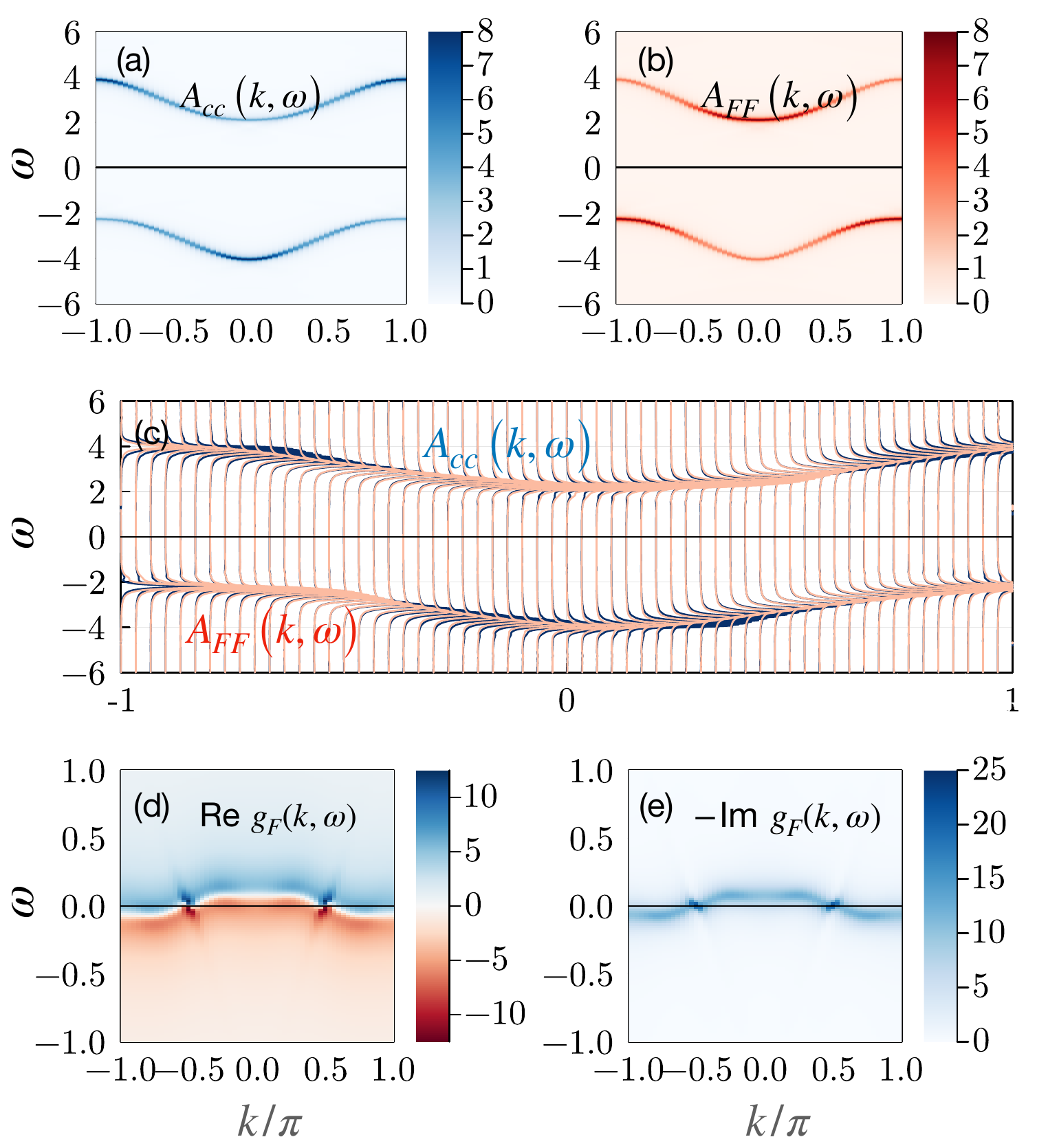}
    \caption{\small Numerical results for the spectral function for $J/t_c=2$. (a) The conduction electron component of spectral function (b) The composite f fermion component of spectral function. (c) The line plots of both c(blue) and f(red) fermion spectral functions. (d) and (e) The real and imaginary part of $g_F (k,\omega)$ defined in Eq.\,\pref{eq:paneld} }
    \label{fig:fig2}
\end{figure}

\begin{figure}[b]
    \includegraphics[width=1\linewidth]{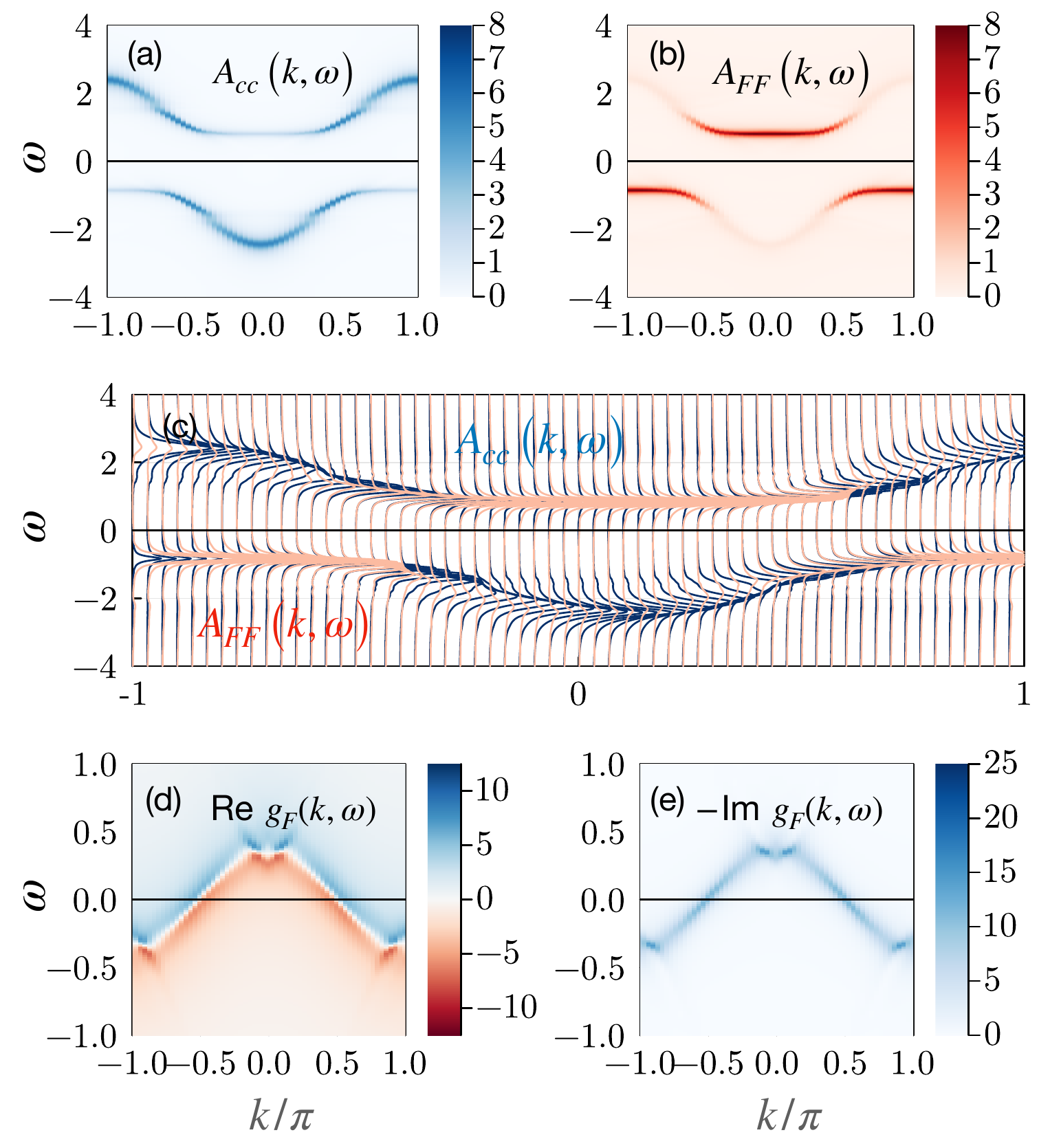}
    \caption{\small Numerical results for the spectral function for $J/t_c=0.9$. (a) The conduction electron component of spectral function. (b) The composite f fermion component of spectral function. (c) The line plots of both c(blue) and f(red) fermion spectra functions. (d) and (e) The real and imaginary part of $g_F (k,\omega)$ defined in Eq.\,\pref{eq:paneld} }
    \label{fig:fig3}
\end{figure}

\subsection{Matrix Spectral Function}

To examine the independence of the $c$ and $F$ fields, it is useful to
combine them into a spinor 
\begin{equation}\label{}
\psi_{\sigma } (x)= \pmat{c_{x\sigma }\cr F_{x\sigma }}, 
\end{equation}
allowing us to define a retarded matrix Green's function 
\begin{eqnarray}\label{}
[{\cal G} (x_{1},x_{2},t)]_{\alpha\beta }&=& -i\theta(t)\langle \{\psi_{\alpha \sigma } (x_{1},t),\psi \dg_{\beta \sigma } (x_{2},0)\}\rangle\cr 
 &\equiv & 
\begin{pmatrix}
G_{ cc} & G_{ cF}   \\
G_{ Fc} & G_{FF}
\end{pmatrix},
\label{eq:green}
\end{eqnarray}
where $\theta(t)$ is the step function. ${\cal G}$ defines a matrix of 
amplitudes for the $c$ and $F$ fields. The $G_{cF}$ component 
\begin{eqnarray}
G^R_{cF}(x_1,x_2,t)  &=& -i \,\theta(t) \braket{ \{c_{x_1}(t) , F_{x_2}^\dag (0) \} }, \label{eq:g_entry}
\end{eqnarray}
determines the amplitude for a composite  $F$ to convert to a
conduction electron. 

We are primarily interested in the properties of a  translationally invariant Kondo lattice, with  momentum-space Green's function
\bea
{\cal G}(k,t) &=& - i \theta(t)\langle  \{ \psi_{k\sigma}(t), \psi\dg_{k\sigma}(0)\}\rangle\cr
&=& \frac{1}{L}\sum_{i,j}e^{ik(x_i-x_j)}{\cal G}(x_i,x_j;t)
\eea
In our numerical calculations, we estimate this Green's function using the expression for a translationally invariant system simply applied to finite size Green's function ${\cal G}(x_i,x_j;t)$. We then perform a discrete Fourier transform on ${\cal G}$ to obtain 
\be
{\cal G}(k,\omega) = \sum_{j=1,N_t}\Delta t \ e^{i \omega t_j}{\cal G}(k,t_j),
\ee
where $\Delta t = T/N_t$ is the spacing of the $N_t$ time-slices over the total duration $T$ of the time evolution, $t_j=j \Delta t$ and the frequencies are sampled at the values $\omega_n= {2\pi}n/{T}$. 

Although we independently compute the four components of ${\cal G}(k,\omega)$, the kinematics of the Kondo lattice imply that  they are not independent, which provides us a means to test and interprete our calculations.   From the Heisenberg equations of motion of the conduction electron operators in the translationally invariant limit,  $i \partial_t c_{k\alpha} = [c_{k\alpha},H]$, 
\be i \partial_t c_{k\alpha} =   \epsilon_c(k) c_{k\alpha} + (3J/2) F_{k\alpha},
\ee
where $\epsilon_c(k) = -2 t \cos(k) $ is the dispersion of the conduction electrons and $F_{k\alpha} = L^{-1/2}\sum_x e^{-i kx}F_{x\alpha}$ is the Fourier transform of the composite fermion.   It follows that  
\bea
[ i \partial_t - \epsilon_c(k)]G_{cc}(k,t) &=&  (3J/2) G_{Fc}(k,t) + \delta(t),\cr
[ i \partial_t - \epsilon_c(k)]G_{Fc}(k,t) &=&  (3J/2) G_{FF}(k,t).
\eea
When we transform these equations into the frequency domain, replacing $i\partial_t \rightarrow z=\omega+i\eta$,
we see that $G_{cc}$ and $G_{cF}$ are entirely determined in terms of $G_{FF},
$\bea\label{relations}
G_{cc} &=& g_c  + g_c (3J/2) G_{FF}(3J/2) g_c\cr 
G_{cF} &=& G_{Fc}= g_c(3J/2)  G_{FF}
\eea
where we have suppressed the $(k,z)$ label on the propagators, and $g_c = [z- \epsilon_c(k)]^{-1}$ is the bare conduction electron propagator.  Although these equations closely resemble the Green's functions of a hybridized Anderson model, with hybridization $3J/2$, we note that $G_{FF}$ represents a composite fermion. 

From these results, it follows that without any approximation,  the inverse matrix Green's function can be written in the form 
\be\label{matrixG}
{\cal G} ^{-1}(k,z) = \begin{pmatrix}
z - \epsilon_{c}(k) & -3J/2 \cr
-3 J/2 &  g^{-1}_{F}(k,z)
\end{pmatrix}
\ee
where $g_{F}(k,z)$ is  the one-particle irreducible composite Green's function, determined by
\be
g_F(k,z)=\Big[\frac{1}{G_{FF}(k,z)}+\frac{(3J/2)^2}{z-\eps_c(k)}\Big]^{-1}.
\ee
This quantity corresponds to the unhybridized composite fermion propagator.  By reinverting \eqref{matrixG} we can express the original Green's functions in terms of $g_F(k,z)$ as follows
\bea
G_{cc}(k,z)&=& \frac{1}{z - \epsilon_c(k)- (3J/2)^2 g_F(k,z)}\cr
G_{FF}(k,z)&=& \frac{1}{g_F^{-1}(k,z)- \frac{(3J/2)^2 }{z- \epsilon_c(k)}}.
\eea
These are exact results, which even hold for a ferromagnetic, $J<0$, Kondo lattice.  By calculating ${\cal G}$ and inverting it, we can thus check the accuracy of our calculation, and we can extract the irreducible $F$ propagator $g_F(k,z)$. 

From this discussion, we see that the ${\cal G}^R$ matrix offers information about both the individual excitations and their hybridization. If the Kondo effect takes place, i.e if $J>0$ is antiferromagnetic, then we expect the formation of an enlarged Fermi surface, driven by the formation of sharp poles in the composite fermion propagator $g_F$. For example, in the special case where the Green's function $g_F$ develops a sharp quasiparticle pole, then we expect
 $g_F(k,\omega) \sim Z_f/[\omega - \epsilon_f(k)]$, allowing us to identify $V = Z (3J/2)$ as an emergent hybridization. 

\subsection{Spectral Functions: Numerical Results}


The spectral function is associated with the Green's function by
\begin{eqnarray}
A_{cc}(q,\omega) &=&  -\frac{1}{\pi} \text{Im}\Big[ G^R_{cc}(q,\omega +i \delta) \Big] \\
A_{FF}(q,\omega) &=&  -\frac{1}{\pi} \text{Im}\Big[ G^R_{FF} (q,\omega+i \delta) \Big] \ .
\end{eqnarray}  
The set of $(q,\omega)$ values for which the spectral function has a maximum is the analogue of a 
band structure for an interacting system. We show the spectral functions computing using MPS
for the cases of \mbox{$J/t_c = 2$} and \mbox{$J/t_c=0.9$} in Fig.~\ref{fig:fig2} and Fig.~\ref{fig:fig3} respectively.

 Figs. 3(e) and 4(e) show the quantity 
\begin{equation}
\text{Im} [g_F(k,\omega+i\eta)]= \text{Im} \frac{1}{ [({\cal G}^R)^{-1}(\omega+i\eta)]_{FF}} \label{eq:paneld}
\end{equation}
where the denominator is $(2,2)$ entry of 2-by-2 matrix $({\cal G}^R)^{-1}$. This quantity can be interpreted as the Green's function of the unhybridized $F$ electrons.

The most striking feature of spectral functions in Figs. \ref{fig:fig2} and \ref{fig:fig3} is the sharp and narrow bands indicative of long-lived and dispersing quasiparticles. For the larger Kondo coupling of $J/t_c=2$, the spectra consist of two cosine dispersion curves $-\cos(k)\pm \Delta E/2$ shifted to positive and negative frequencies. For the smaller Kondo coupling of $J/=0.9$, the dispersion can be thought to arise as the hybridization of a dispersing band (mostly $c$ content) and a localized band (mostly $F$ content). 
It is apparent that in both cases, the dispersion can be approximately reproduced using a two-band fermionic model. Assuming that this is so, the quantity $\text{Im}[g_F(k,\omega-i\eta)]$ shown in
panels \ref{fig:fig2}(d) and \ref{fig:fig3}(d) can be interpreted as the bare dispersion the putative $F$ fermion would need to have in order to reproduce the observed spectral functions. In both cases, a non-zero dispersion is discernible which is more significant in the $J/t_c=0.9$ case. 
Since in absence of Kondo interaction, composite fermions are localized $\braket{F\dn_iF\dg_j}\propto\delta_{ij}$, this bare dispersion is naturally associated with dynamically generated magnetic coupling between the spins due to RKKY interaction which gives rise to dispersing spinons.

\subsection{Comparison with strong coupling and mean-field}
It is natural to expect some of the numerical results to match those obtained in the strong Kondo coupling limit $J_K/t_c\gg 1$. When $t_c=0$ the decoupled sites each have the spectrum
\be
H=\left\{\bmx{lll}J/2, \qquad\quad & S=1, & n=1, \\ 0, & S=1/2, & n=0,2 \\ -3J/2, & S=0, & n=1.\emx\right.
\ee
where the quantum numbers $S$ and $n$ are the total spin and charge at that particular site. 
Creating or annihilating a particle from the ground state has the energy cost of $E_1=3J/2$. To understand how the ground state and single-particle excited states evolve for a finite $t$, we have carried out a perturbative analysis for the full $2\times 2$ Green's function in the Appendix \ref{sec:scsingle} and found that to lowest orders in $t_c/J$,
\be
{\cal G}^{-1}(k,z)=z\bb 1-H, \qquad H=\mat{\eps_k & V \\ V & 0},
\ee
where $V=E_1$. The eigenenergies are
\be
E_\pm(k)=\frac{1}{2}\Big[\eps_k\pm\sqrt{\eps_k^2+V^2}\Big],\label{eq2band}
\ee
which confirms the picture of two hybridized bands. Here $z$ is the complex frequency and $\eps_k=-2t_c\cos k$ is the bare dispersion of the conduction electrons. Note that to this order, the dispersion of the bare $F$ band is not captured in agreement with previous results \cite{Trebst06}.

\begin{figure}[h!]
\includegraphics[width=0.49\linewidth]{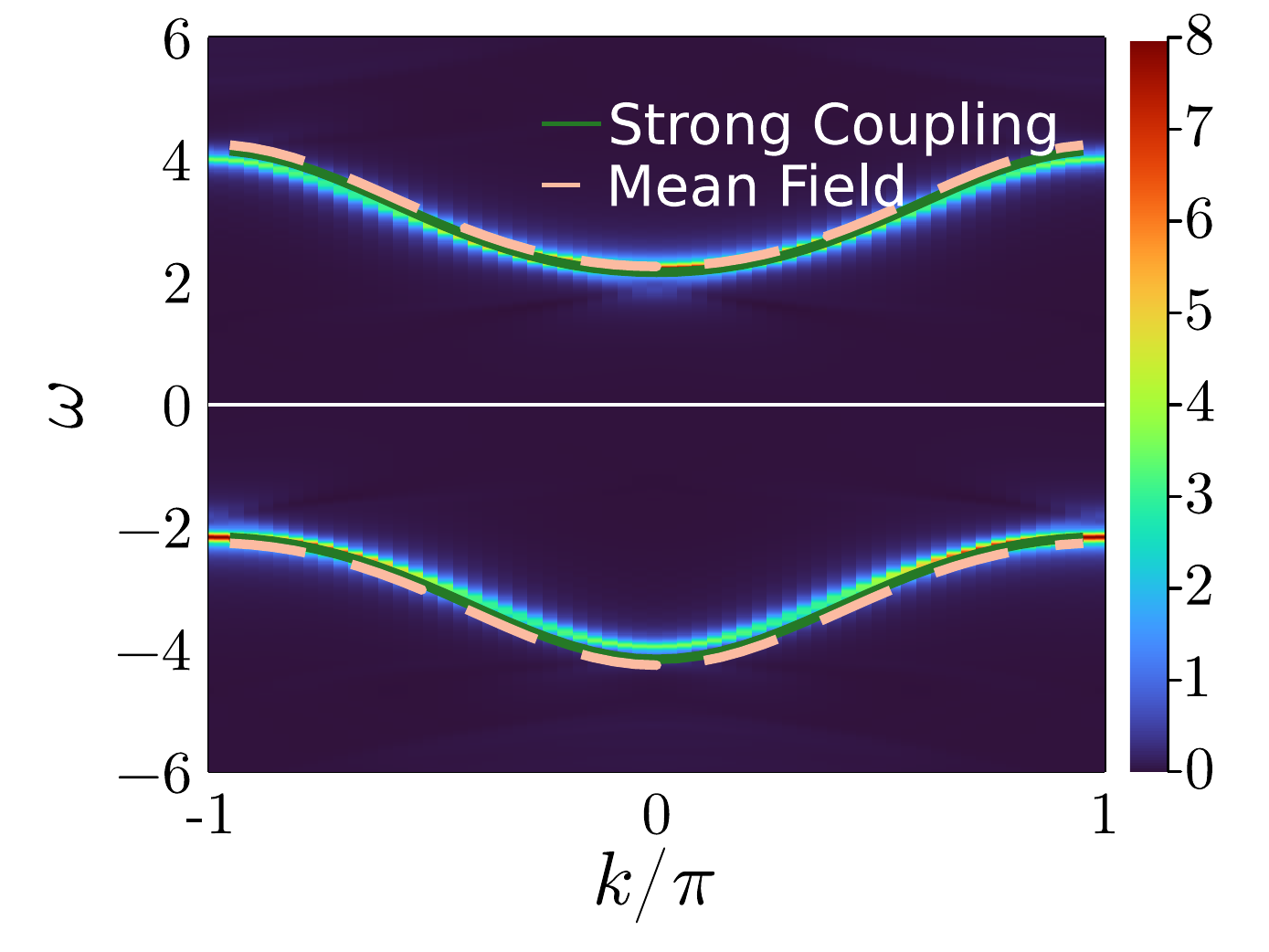}
\includegraphics[width=0.49\linewidth]{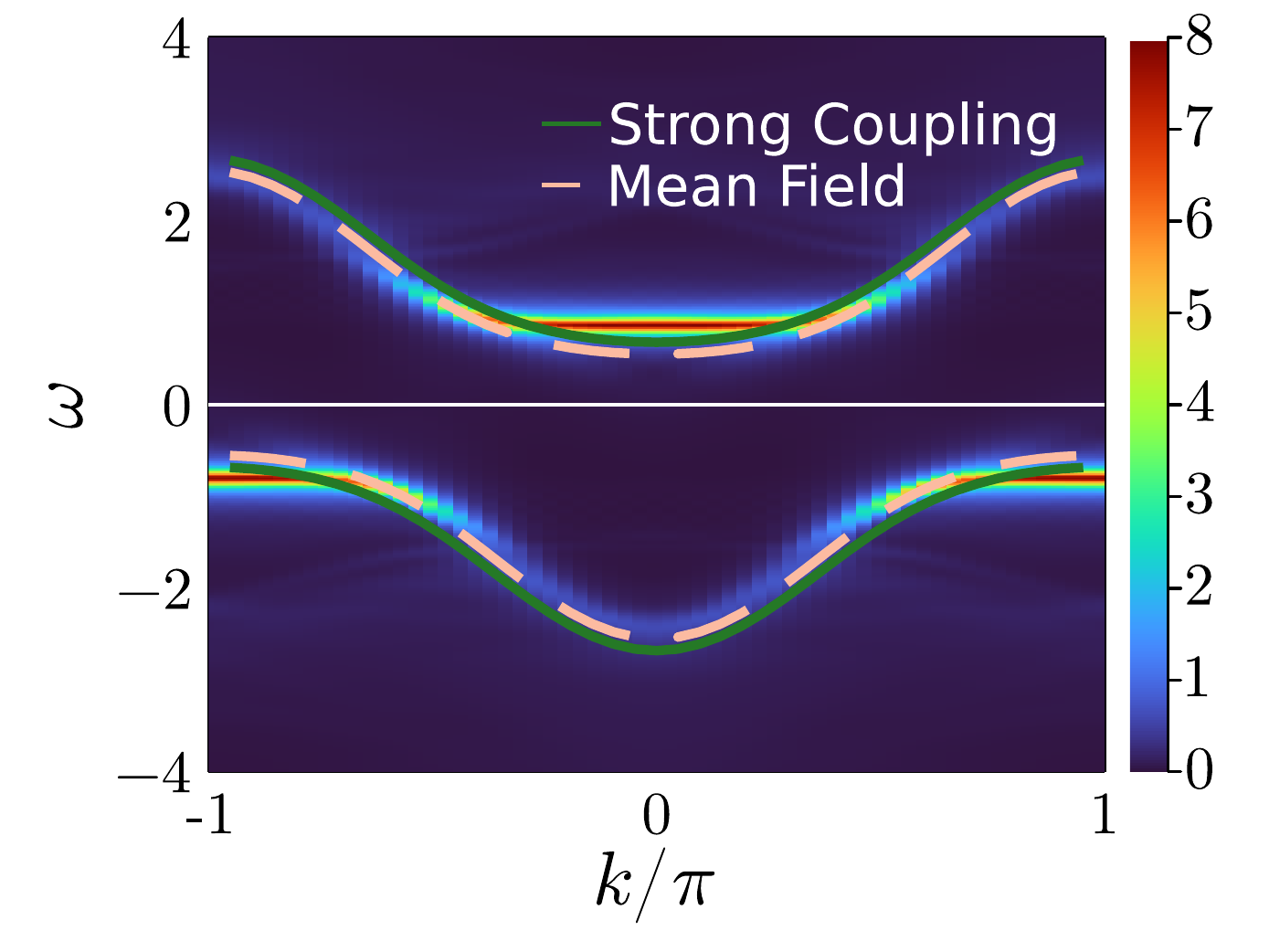}
\caption{\small A comparison of $A_{FF}(k,\omega)$ with the dispersion from mean-field theory (solid line) and strong-coupling expansion (dotted line). Left and right panels are $J/t_c=2$ and $J/t_c=0.9$, respectively. For $J/t_c=0.9$, there is a noticeable deviation of the perturbative or mean-field results from the numerical results near $k=0$ for the upper band and near $k=\pi$ for the lower band.}\label{fig:comp}
\end{figure}

The quasi-particle spectrum \pref{eq2band} has the same form as in the large-N mean-field theory, with the difference that the value of $V$ is determined from self-consistent mean-field equation [see Appendix \ref{sec:mft}]. We have plotted $A_{FF}(k,\omega)$ spectra in Fig.\,\pref{fig:comp} along with predictions from strong-coupling expansion and mean-field theory. Overall, a good agreement is found albeit deviations start to appear at lower Kondo coupling of $J=0.9$.

One artifact of the mean-field theory is that the hybridization $V$ is systematically underestimated which can be traced back to the relation between $F$ and $f$ in Eq.\,\pref{eq2} and \pref{eq10}. For example, at the strong coupling limit, the mean-field theory predicts $V=J/2$ [see Appendix \ref{sec:mft}]. In order to get an agreement, we had to re-scale $V\to 3V/2$ when comparing mean-field results to numerical results on the interacting system. One can alternatively motivate this rescaling by viewing the mean-field theory as an effective model, where the $V$ parameter in the mean-field is ``renormalized'' from the bare hybridization. 

\subsection{Evolution of Single Particle States} 

A vivid demonstration of the particle nature of $F$ excitations can be seen by a calculating the motion of a composite fermion wavepacket. Here, it proves useful to take account of the spatially dependent normalization of the composite fermions, defining a normalized composite fermion as follows
\begin{equation}
    f_{x\sigma}= \frac{1}{\sqrt{Z(x)}}F_{x\sigma},
\end{equation} 
where the normalization, is calculated from the measured expectation value of the anticommutator
\begin{equation}
Z(x) = \langle {\rm GS}\vert \{F_{x\sigma},F\dg_{x\sigma}\}\vert {\rm GS}\rangle = 2\langle {\rm GS} \vert F_{n\sigma}F\dg_{n\sigma}\vert{\rm GS}\rangle.
\end{equation}
Here the second expression follows from particle-hole symmetry.  This normalization guarantees that the expectation value of the anti-commutator is normalized
$\langle \{ f_{x_1\sigma}, f\dg_{x_2\sigma' }\}\rangle=\delta_{x_1x_2}\delta_{\sigma\sigma'}$.
In the ground-state, $Z(x)$ is a constant of motion, which with our definition of $F_{x\sigma}$ \eqref{eqn:F}, is unity in the strong-coupling limit. However, at intermediate coupling, $Z(x)$  becomes spatially dependent near the edge of the chain. 

Consider a  wave-packet 
\begin{equation}
\vert w\rangle = \sum_{n}\phi_f(x_n)f\dg_{x_n\sigma}\vert {\rm GS}\rangle
\end{equation}
where 
\begin{equation}
\phi_{f}(x_n) = \frac{1}{\sqrt{\cal N}}e^{i k_0x_n}e^{-\frac{(x_n-y)^2}{4\sigma^2}}
\end{equation}
is a normalized wave-packet centered at $y$  with momentum $k_0$. The time-evolution of this one-particle-state will give rise to a 
state of the form
\begin{eqnarray}
\vert w(t)\rangle &= &e^{-iHt}\vert w \rangle \cr 
&=& \sum_{n}\bigl[\phi_f(x_{n},t)f\dg_{x_{n}\sigma}+\phi_c(x_{n},t)c\dg_{x_n\sigma}\bigr]\ket{\rm GS} \cr &+& \dots
\end{eqnarray}
where the $\dots$ denotes the many-particle states that lie outside the Hilbert space of one conduction and one composite fermion. Taking the overlap with the states $f\dg_{x_n\sigma}\vert {\rm GS}\rangle$ and $c\dg_{x_n\sigma}\vert {\rm GS}\rangle$,  the coefficients of the wavepacket can be directly related to the Green's functions as follows
\begin{eqnarray}
    \phi_f(x_n,t) &=& \langle{\rm GS} \vert f_{x_n\sigma}e^{-i H t} f\dg_{x_{n'}\sigma }\vert {\rm GS}\rangle \cr &=&i e^{-iE_g t}\sum_{n'}{\cal G}^>_{ff}(x_n,x_{n'};t)\phi_f(x_{n'}),
\end{eqnarray}
and similarly 
\begin{equation}
       \phi_{c}(x_n,t) =ie^{-iE_g t}\sum_{n'}{\cal G}^>_{cf}(x_n,x_{n'};t)\phi_f(x_{n'}),
\end{equation}
where $E_g$ is the ground-state energy and
\begin{eqnarray}
{\cal G}^>_{ff}(x_n,x_{n'};t)&=& -i \langle f_{x\sigma}(t)f
\dg_{x'\sigma}(0)\rangle,\cr
G^>_{cf}(x_n,x_{n'};t)&=& -i \langle c_{x\sigma}(t)f
\dg_{x'\sigma}(0)\rangle,
\end{eqnarray}
Using the Green's functions computed from the MPS time-evolution, we can thus evaluate the time-evolution of the wave-packet. 

\begin{figure}
\includegraphics[width=1\linewidth]{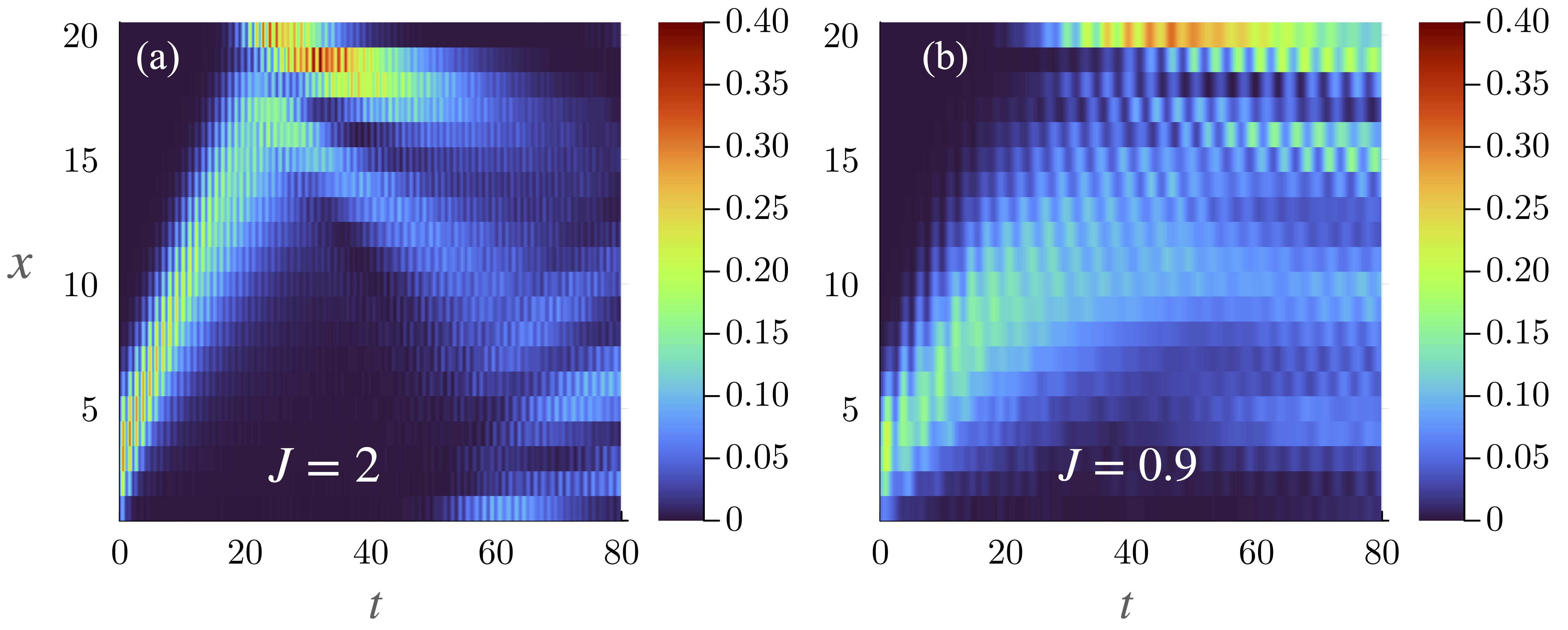}  
\caption{\small Evolution of a composite fermion wave-packet for (a) $J=2.0$ and (b) $J=0.9$. The initial wavefunction is Gaussian with $\sigma^2=7$ and momentum $k_0=1.4$, in units where the lattice spacing $a=1$ is unity.}\label{fig:figWP}
\end{figure}

 Figure~\ref{fig:figWP} shows the evolution of the probability density $|\phi_f(x,t)|^2+|\phi_c(x,t)|^2$ of an initial Gausian wavepacket for two values of $J/t=2$ and $J/t=0.9$. In the former case the composite fermion wavepacket moves ballistically until it is scattered by the boundary of the system. In the $J/t=0.9$ case, however, the wave-packet undergoes significant dispersion and decay with distance, appearing to "bounce" long before reaching the wall.  One possible origin of this effect, is the break-down of the Kondo effect in the vicinity of the wall, due to a longer Kondo screening length $\xi = v_F/T_K$. 

\subsection{Interpretation of single-particle results}


One of the most remarkable aspects of this comparison, is the qualitative agreement between the spectral functions derived from the matrix product, strong coupling and large $N$ expansion. 
In all three methods, we see that the description of the spectral function requires a two-band description.  Our matrix product simulation shows that the composite fermion propagator $g_F$ contains sharp poles at $k=\pm \pi/2$, $\omega=0$, which reflect a formation of composite fermion bound-states, as if the $F$ fields behave as sharp bound-states. 

The single-particle excitation spectrum exhibits a coherent two-band fermionic model which continues to low $J/t_c$. This suggests that the composite $F$-excitations, behave as bound-states of conduction electrons and spin flips of the local moments, forming an emergent Fock  space that is effectively orthogonal to that of the conduction electrons, so that $c$ and $F$ fermions are effectively independent fields. In effect, the microscopic Hilbert-space of the spin degrees of freedom has morphed into the Fock-space of the F-electrons.

In the large N limit, the composite fermion F is synonymous with a fractionalization of the local moments into half-integer spin fermions, moving under the influence of an emergent $U(1)$  gauge field that imposes the constraints.
From the single-particle excitation spectrum alone, aside from hybridization with conduction electrons,  these emergent $F$ fermions appear to be free excitations: the comparison with mean-field theory suggests that the original spin is fractionalized to $\vec S_F=F\dg\frac{\vec\sigma}{2}F$. As shown in \ref{secFcom}, in the strong coupling regime $\vec S_F\sim\frac{1}{3}\vec S$.

How accurate and useful is this picture? If $F$ electrons are indeed free beyond one-particle level, their higher-order Green's functions (including two-point functions of $\vec S$) would factorize into spin-1/2 fermions. We now investigate this possibility.


\section{Composite Fermions: Two-Particle Properties and Spin Susceptibility}
\begin{figure}
    \includegraphics[width=1\linewidth]{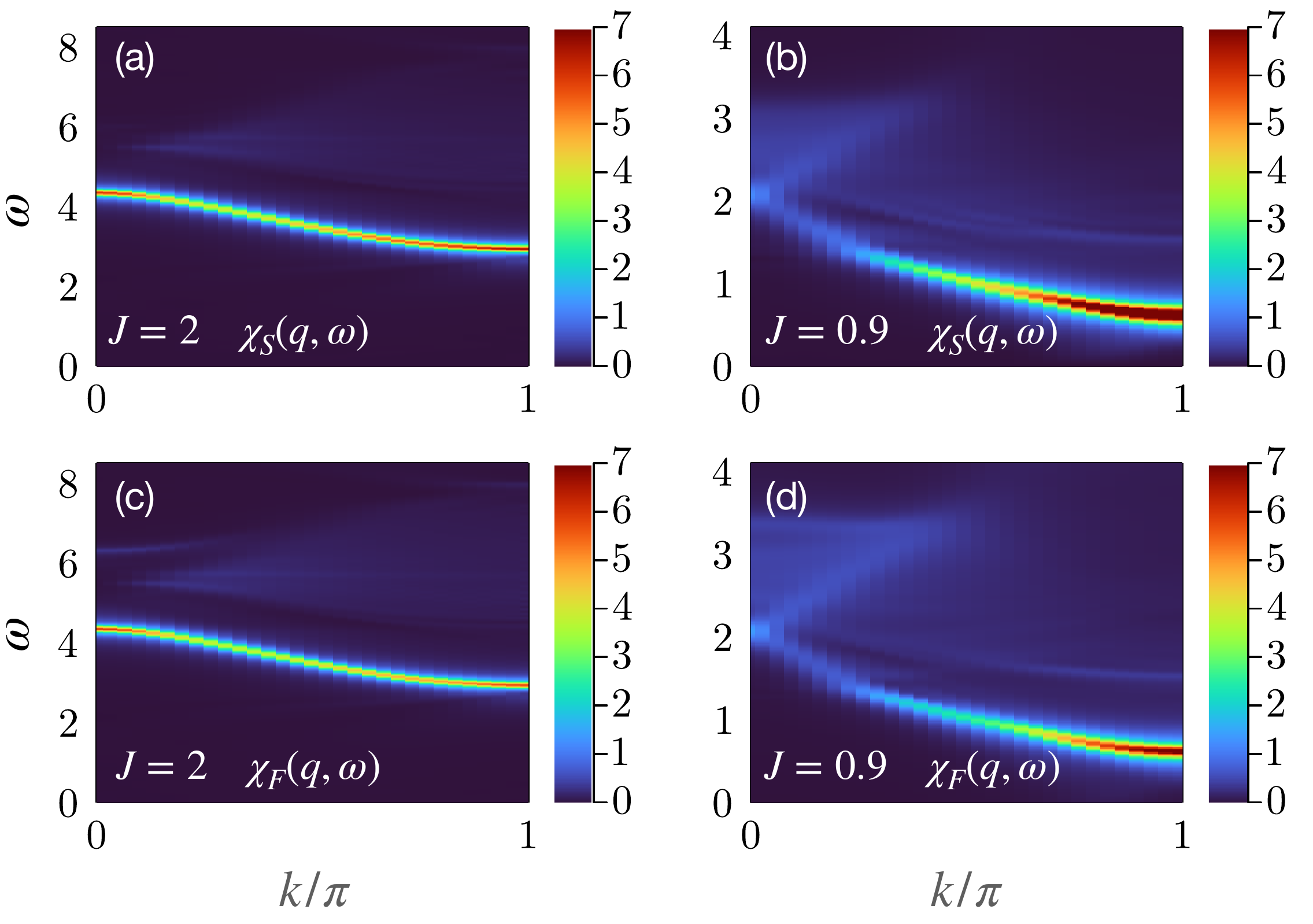}
    \caption{\small (a,b) The spin susceptibility $\chi_S(q,\omega)$ and (c,d) the composite fermion susceptibility $9\chi_F(q,\omega)$ for two values of $J/t_c=1.8$ and $J/t_c=4$ case. The two are nearly identical with minor differences at small momenta and high frequency.} 
    \label{fig:Sqw}
\end{figure}

Next, we turn to the two-particle spectrum and focus on the spin susceptibility
\bea
\hspace{-.5cm}\chi_S(q,\omega)&=&\sum_{n,m}\int_0^\infty dt\ {e^{i[(\omega+i\eta)t-q(n-m)]}}\chi(x_n,x_m,t)\cr
\chi(x_1,x_2,t)&=&
-i\braket{[S^-(x_1,t),S^+(x_2,0)]}.\label{eqchi_S}
\eea
which can be probed experimentally. This function satisfies the sum-rule $\int{{d\omega}}\chi''_S(q,\omega)={2\pi}\braket{S^z}=0$ for any $q$. 
Fig.\,\ref{fig:Sqw} shows $\chi''_S(q,\omega)$ 
for two values of $J/t$, computed from the Fourier transform of $\chi  (x_1,x_2,t)$ and using the same Fourier transform procedure as in  \eq{eq:fourier}. 
A broad incoherent region and at least one sharp dispersing mode (at low positive frequency) is visible. The latter is more pronounced at higher $J/t=4$  compared to $J/t=1.8$.
A spin-flip creates a localized triplet. Since only the total magnetization is conserved, the triplet can move in the lattice forming a coherent magnon band. However, in this interacting system, the magnon can decay into many-body states and the reduced weight of the coherent band is compensated by the incoherent portion of the spectrum. 

In the previous section, based on the behavior of $F$ particles we conjectured that the spin $\vec S$ is proportional to $\vec S_F=\frac{1}{2}F\dg \vec\sigma F$. The relationship $\vec S_F= \frac{1}{3}\vec S$ is in fact correct in the strong Kondo coupling limit (appendix \ref{secFcom}). To test its validity beyond this limit, we compare $\chi_S(q,\omega)$ with $9\chi_F(q,\omega)$ defined in terms of composite fermions:
\bea
\hspace{-.5cm}\chi_F(q,\omega)&\equiv&-i\sum_{n,m}\int_0^\infty dt\ {e^{i[\omega+i\eta)t-q(n-m)]}}\chi_F(x_n,x_m,t)\cr
\chi_F(x_n,x_m,t)&=&
i\braket{[S_F^-(n,t),S_F^+(m,0)]}.
\eea
and involves four-point functions like
$
\braket{ F^\dag_{n\downarrow}(t) F_{n\uparrow}(t) F^\dag_{0\uparrow}(0) F_{0\downarrow}(0) }
$.
We see that the two are exactly equal, proving the relation $\vec S\sim 3\vec S_F$ at least within the two-particle sector.

\begin{figure}
    \centering.    
    \includegraphics[width=1\linewidth]{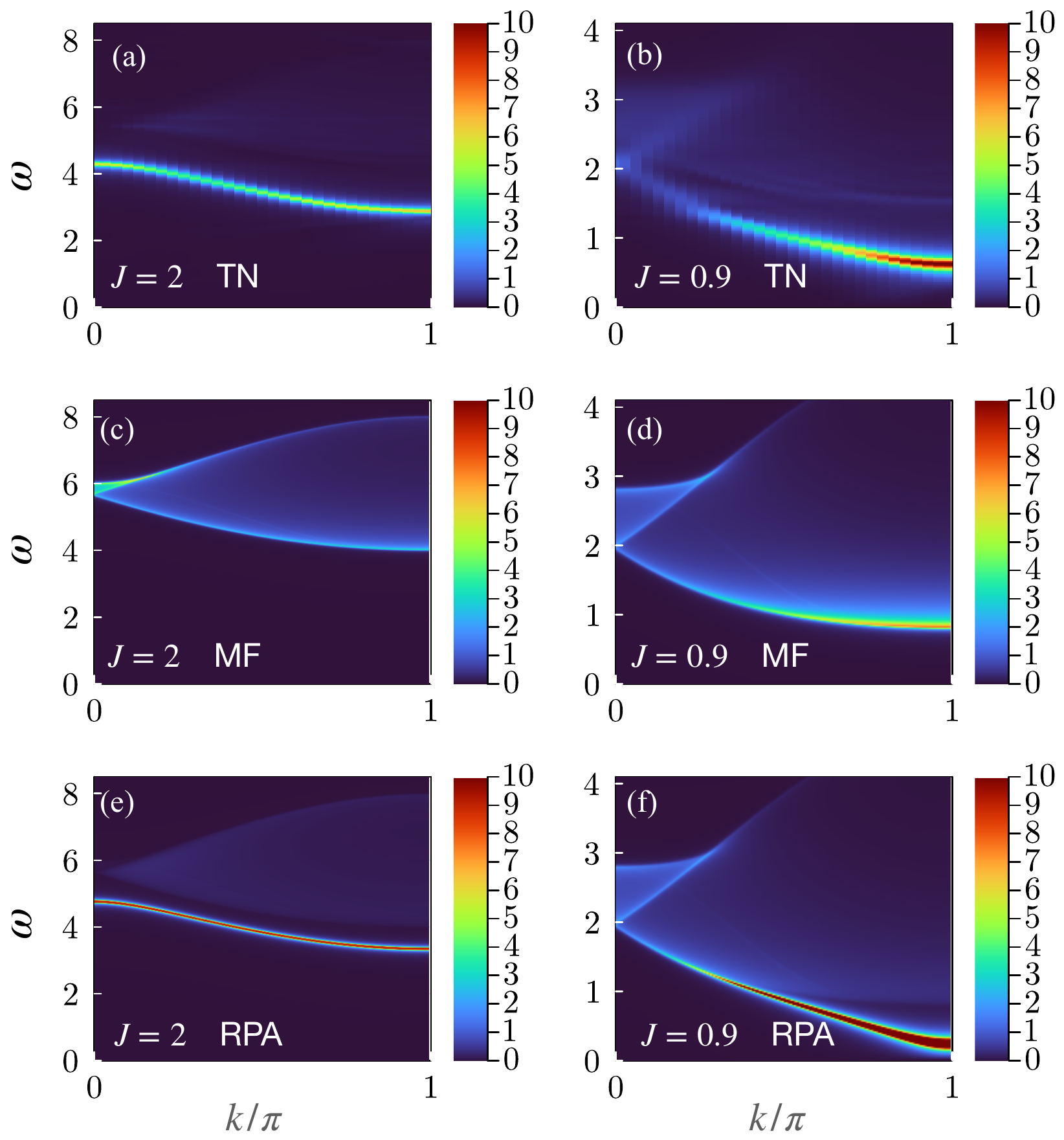}
    \caption{Spin susceptibility $\chi_S(q,\omega)$. The first rwo shows tensor network results for (a) $J/t=2$ and (b) $J/t=0.9$. This is compared with large-N mean-field theory (MF) results in (c) $J/t=2$ and (d) $J/t=0.9$, and random-phase approximation (RPA) results in (e) $J/t=2$ and (f) $J/t=0.9$. The parameters in (e) and (f) are $U'/t=-2$ and $U'/t=-0.75\sin(q/2)$. A generally $q$-dependent interaction between quasi-particles within RPA captures both magnon branch and the details of the correlation function at $q\sim 0$.}
    \label{fig:TN_MF_compare}
\end{figure}

However, while this relation seem to hold,  fractionalization as seen in 1D Heisenberg AFM requires the four-point function $\chi_F$ to be expressible in terms of the convolution of two single-particle propagators. To examine this possibility, we compare the
spin susceptibility $\chi''_S(q,\omega)$ with the mean-field spin susceptibility $\chi''_{MF}(q,\omega)$ computed from the convolution of two f-electron propagators Fig. \ref{fig:TN_MF_compare}. The mean-field dynamical susceptibility contains a continuum of excitations bordered by two sharp lines that result from the indirect gap between the f-valence and f-conduction bands (lower sharp line) and the c-valence and c-conduction bands (upper sharp line) of the fractionalized Kondo insulator.  A particularly marked aspect of the mean-field description in terms of fractionalized f-electrons is the continuum at $q\sim 0$ which stretches from the hybridization gap ($2V$) out to the half band-width of the conduction band. At finite $q$ this continuum evolves into a characteristic inverted triangle-shaped continuum. 
At strong coupling, $J/t=2$ $\chi''_S(q,\omega)$ contains a sharp magnon peak, and the triangle-shaped continuum is absent. This is clearly different from $\chi''_{MF}(q,\omega)$. However at weaker coupling $J/t=0.9$, the MPS susceptibility is qualitatively similar to the mean-field theory, displaying the triangle-shaped continuum around $q\sim 0$ and a broadened low energy feature that we can associate with the indirect band-gap excitations of the f-electrons. It thus appears that at strong-coupling, the f-electrons are confined into magnons, whereas at weak-coupling the spins have fractionalized into heavy fermions. 


\subsection{Strong coupling and mean-field perspective}
To gain further insight into the dynamical spin susceptibility, we discuss the two-particle sector from both strong coupling and mean-field perspectives. It is useful to generalize the Hamiltonian of Eq.\,\pref{eqn:H} by including a Coulomb repulsion $U>0$, i.e. 
\be
H_{\rm generalized}=H+
U\sum_j(c\dg_{j\ua}c\dn_{j\ua}+c\dg_{j\da}c\dn_{j\da}-1)^2
\ee
which favors one electron per site. We assume $U$ is small enough so that the ground state is smoothly connected to the original problem with $U=0$. 

\begin{figure}
    \centering
    \includegraphics[width=1\linewidth]{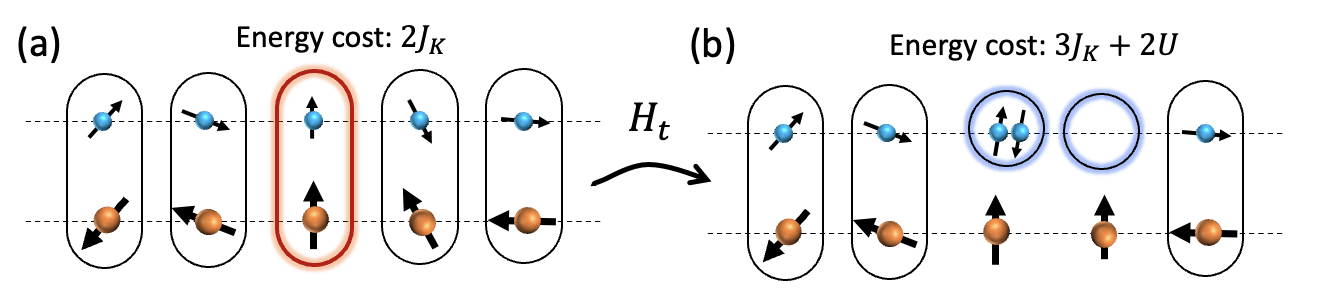}
    \caption{\small The spin-flip is equivalent to creation of a triplet doublon-holon pair at the same site. The pair can move together as a magnon or decay into fractionalized doublon and holon. The former has lower energy if $U'=J_K+2U>0$.}
    \label{fig:holondoublon}
\end{figure}

The starting point is that at the strong coupling, all sites are singlets, and therefore the relation
\be
(\vec S+c\dg\frac{\vec\sigma}{2} c)\ket{K_j}=0
\ee
holds. This means that $\vec S$ can be replaced by  $-\frac{1}{2}c\dg\vec\sigma c$ in $\chi_S$ defined in Eq.\,\pref{eqchi_S}, creating the following strong-coupling picture: A $S^+$ spin-flip can be considered as a creation of a local doublon-holon spin-triplet $T^+$ pair at the same site. Such a state has energies around $E_2=2E_1$ as shown in Fig.,\,\ref{fig:holondoublon}. Under time-evolution, the doublon and holon can move around and recombine at site $n$ where the $T+$ triplet is annihilated. Such a $T^+$ triplet is described by
\be
\ket{F^+}=\sum_{n_1n_2}\psi^+(n_1,n_2)c\dg_{n_1\ua}c\dn_{n_2\da}\ket{\Omega}.
\ee
Including the $U$ interaction, each holon or doublon costs an energy  $E_1+U/2$ and $E_2\to 2E_1+U$.
By acting on this with the Hamiltonian $H\ket{F^+}=E\ket{F^+}$ and projecting the result to within the two-particle excitations, we find that the wavefunction $\psi(n_1,n_2)$ obeys the first-quantized Schr\"odinger equation
\bea
&&\frac{t}{2}\Big[\psi(n_1+1,n_2)+\psi(n_1-1,n_2)-\psi(n_1,n_2+1)\\
&&-\psi(n_1,n_2-1)]+U'\delta_{n_1,n_2}\psi(n_1,n_2)=(E-E_2)\psi(n_1,n_2).\nonumber
\eea
This is a two-particle problem, where the particles interact via the $U'=-J_K-U$ term. Note that a repulsive/attractive interaction among electrons is an attractive/repulsive interaction among doublon and holon. 

In the usual regime ($U\ge 0$) the interaction $U'<0$ is \emph{attractive}. While a continuum of excited states exists, the ground state is a stable magnon boundstate between doublon and holon, with a correlation length that diverges as $U'\to 0$. The continuum is essentially a fractionalized magnon into doublon and holon pairs as can be seen in the $U'=0$ case. It is natural to expect that due to interactions not considered, the doublon-holon pair decay into the ground state. For an attractive $U\le -J_K/2$, the interaction between doublon and holon $U'>0$ is \emph{repulsive}, rendering the bound-state highly excited and unstable.

The eigenstates $\ket{F^+}$ can be used to compute the spin-susceptibility $\chi_S$. The result is shown in Appendix \ref{sec:twopsc}. The result at the strong coupling $J/t=2$ contains a magnon band in good agreement with MPS results. This indicates that while a spin-flip has fractionalized into a doublon-holon pair, there are residual attractive forces in a Kondo insulator that bind the two. 

On the other hand, at the weak coupling $J/t=0.9$ limit, the strong-coupling analysis is incapable of reproducing MPS results around $q\sim 0$. This suggests that $U'$ renormalizes to zero in the small momentum limit. As seen in Fig.\,\ref{fig:TN_MF_compare}(b) in the weak-coupling limit, the strong magnon-like resonance in the MPS dynamical spin susceptibility, broadens and merges with the triangular feature around $q\sim 0$, in sharp contrast to the strong coupling results and more closely resembling the mean-field theory [Fig.\,\ref{fig:TN_MF_compare}(d)]. 

To capture the doublon-holon interaction and the magnion band, the mean-field theory be improved by including a  momentum-dependent residual interaction $U'(q)$ between f-electron quasi-particles within a random-phase-approximation (RPA) framework. The resulting susceptibility can be written as
\be
\chi_q^{\rm RPA}(\omega+i\eta)=\frac{1}{[\chi_q^{\rm MF}(\omega+i\eta)]^{-1}-U'(q)}
\ee
The RPA results is shown in Fig\,\ref{fig:TN_MF_compare}(e,f) for a constant $U'$ in the strong coupling $J/t=2$ case and $U'(q)\sim q$ in the weak-coupling $J/t=0.9$ case, both in good agreement with the MPS results.

Overall, these results indicate that the low-lying spin-1 charge-neutral excitation of the ground state can be regarded as fractionalized into spin-1/2 charge-e single particle excitations that have some residual attraction in one-dimension, forming a magnon branch in the dynamical spin-susceptibility. The disappearance of this magnon branch at $q\to 0$ in the weak-coupling regime and its comparison with RPA suggests that at long distances the residual interaction disappears, leading to deconfined quasi-particles.


\section{Conclusion}

By contrasting strong coupling, mean-field theory and matrix product calculations of the dynamics of the one dimensional Kondo insulator, we  gain an important new perspective into the nature of the excitations in this model. 
There are a number of key insights that arise from our results. 

Firstly, we have been able to show that the composite fermion, formed between the conduction electrons and localized moments behaves as an independent fermionic excitation, giving rise to a two-band spectrum of charge $e$, spin-$1/2$ excitations, with hybridization between the electrons and the independent, composite fermions. Our results are remarkably consistent with the mean-field treatment of the Kondo insulator.  

By contrast, our examination of the dynamical spin susceptibility paints a more nuanced picture of the multi-particle excitations.  At strong-coupling, we can explicitly see that the  triplet holon and doublon combination created by a single spin-flip form a bound magnon, giving rise to a single magnon state in the measured dynamical susceptibilty. Thus at strong coupling, the spin excitation spectrum shows no sign of fractionalization. On the other hand, it can be easily checked that spin-singlet charge-2e excitations are always deconfined. Essentially, two dobulons (or two holons) can never occupy the same site, very much as same-spin electrons avoid each other due to Pauli exclusion, and thus do not interact.

However, at weaker coupling, the dynamical susceptibility calculated using MPS methods, displays a dramatic continuum of triplet excitations  with an inverted triangle feature at low momentum, characteristic of the direct band-gap excitations across a hybridized band of conduction and f-electrons, and high momentum feature that resembles the indirect band-gap excitations of heavy f-electrons. These results provide clear evidence in support of a fractionalization picture of the 1D Kondo insulator at weak coupling. Based on these results, it is tempting to suggest that there are two limiting phases of the 1D Kondo insulator: a strong coupling phase in which the f-electrons are confined into magnons, and a deconfined weak-coupling phase where the local moments have fractionalized into gapped heavy fermions
The emergence of a continuum in the spin-excitation spectrum at weak coupling may  indicate that that the confining doublon-holon interaction at strong coupling, either vanishes, or changes sign at weak coupling, avoiding the formation of magnons.

\subsection{Further Directions}

 It would be very interesting to extend these results to two dimensions.  The strong-coupling analysis of the composite fermion Green's function and the doublon-holon bound-states can be extended to higher dimensions, where it may be possible to calculate a  critical $J$ at which confining doublon-holon bound-state develops. Further insight might be gained into the two-dimensional dimensional Kondo insulator using matrix-product states on Kondo-lattice strips, or alternatively, by using fully two dimensional tensor-network approaches or sign-free Monte-Carlo methods\,\cite{Danu2021}.

\subsection{Discussion: Are heavy fermions in the Kondo lattice fractionalized excitations? }

 The 1D Kondo lattice is the simplest demonstration of Oshikawa's theorem \cite{Oshikawa2000}: namely the expansion of a Fermi surface through spin-entanglement with a conduction electron sea. Traditionally, the expansion of the Fermi surface in the Kondo lattice is understood by regarding the Kondo lattice as the adiabatic continuation of a non-interacting Anderson model from small, to large interaction strength\cite{Martin82}. Yet viewed in their own right, the  ``f-electron" excitations of the Kondo lattice are emergent.   
 
 Our calculations make it eminently clear that in the half-filled 1D Kondo lattice, the f-electrons created by the fields \begin{equation} f\dg_{j\sigma} = \frac{1}{\sqrt{Z(j)} }F\dg_{j\sigma}, \end{equation}  form an emergent Fock space of low energy, charge $e$ excitations that expand the Fermi sea from a metal, to an insulator. Less clear, is the way we should regard these fields from a field-theoretic perspective. From the  large-$N$ expansion it is tempting to regard heavy-fermions as a fractionalization of the localized moments, $\vec S_j\rightarrow f\dg_{j\alpha}\left(\frac{\vec \sigma}{2}\right)_{\beta}f_{j\beta}$.  Our calculations provide support for this picture in the weak-coupling limit of the 1D Kondo lattice, where we see a intrinsic dispersion of the underlying $F$ electrons, reminiscent of a spin liquid, and a continuum of $S=1$ excitations in the dynamical spin susceptibility. 

Yet the use of the term ``fractionalization" in the context of the Kondo lattice is paradoxical, because the excitations so-formed are self-evidently charged. Field-theoretically, the spinons transform into heavy fermions, acquiring electric charge while shedding their gauge charge via an Anderson-Higgs effect that pins the internal spinon and external electromagnetic gauge fields together. 

{Why then, can we not regard the f-electrons of the Kondo lattice as ``Higgsed''-fractionalized excitations?  This is because the classical view of confinement \cite{ShenkerFradkin} argues that confined and Higgs phases are adiabatic limits of single common phase:
i.e. the excitations of a Higgs phase are confined. }
Yet on the other hand, we can clearly see the one and two-particle f-electron excitations, born from the localized moments, not only in the large $N$ field theory, but importantly,  in the matrix product-state calculations of the 1D Kondo lattice.   Moreover, a recent extension of Oshikawa's theorem  extends to all $SU(N)$ Kondo lattices \,\cite{Hazra21}, suggests that the large $N$ picture involving a fractionalization of spins into heavy fermions is a valid  description of the large Fermi surface in the Kondo lattice.  How do we reconcile these two viewpoints? Further work, bringing computational and analytic techniques together, extending our work to higher dimensions will help to clarify these unresolved questions. 

Y.~K. acknowledges discussions with E.~Huecker.  
This research was supported by the U. S. National Science Foundation division of Materials Research, grant DMR-1830707 (P. C. and also, Y. K during the initial stages of the research). 

\appendix

\section{$F$ commutation relations}\label{secFcom}
The composite fermion operators have the expression
\be
F_\alpha=\frac{2}{3}\vec S\cdot\vec\sigma_{\alpha\beta'}c_{\beta'}, \andd F\dg_\beta=\frac{2}{3}c\dg_{\alpha'}\vec\sigma_{\alpha'\beta}\cdot\vec S\label{eq:A1}
\ee
which means
\be
F_\ua=\frac{2}{3}(S^zc_\ua+S^- c_\da),\qquad 
F_\da=\frac{2}{3}(-S^zc_\da+S^+ c_\ua).
\ee
The same factor of $2/3$ appears in strong coupling expansion of multi-channel lattices \cite{Ge2022}. The anti-commutation relations are
\be\label{acomm}
\{F\dn_\alpha,F\dg_\beta\}=\frac{4}{9}\sigma^a_{\alpha\beta'}\sigma^b_{\alpha'\beta}(S^aS^bc\dn_\beta c\dg_{\alpha'}+S^bS^a c\dg_{\alpha'}c\dn_{\beta'})
\ee
We use the identities
\bea
c\dg_{\alpha'}c_{\beta'}&=&\mat{c\dg_\ua c\dn_\ua & c\dg_\ua c\dn_\da\\ c\dg_\da c\dn_\ua & c\dg_\da c\dn_\da }_{\alpha'\beta'}=[\frac{n}{2}\bb 1+\vec s\cdot\vec\sigma^T]_{\alpha'\beta'}\nonumber\\
&=&[\frac{n}{2}\bb 1+\vec s\cdot\vec\sigma]_{\beta'\alpha'},
\eea
and
\be
c_{\beta'}c\dg_{\alpha'}=[(1-\frac{n}{2})\bb 1-\vec s\cdot\vec\sigma]_{\beta'\alpha'},\qquad
\ee
to recast \eqref{acomm} as
\bea
\{F\dn_\alpha,F\dg_\beta\}&=&\frac{4}{9}(\sigma^a\sigma^b)_{\alpha\beta}\Big[S^aS^b\frac{n}{2}+S^bS^a(1-\frac{n}{2})\Big]\nonumber\\
&&\qquad-\frac{4}{9}\Big[[S^a,S^b]s^c(\sigma^a\sigma^c\sigma^b)_{\alpha\beta}\Big]\qquad
\eea
Using
\be
[S^a,S^b]={i}\eps^{abd}S^d, \andd \sigma^a\sigma^a=3\bb 1\label{eqS2}
\ee
we can simplify the anti-commutator to
\bea\label{acomm2}
\{F\dn_\alpha,F\dg_\beta\}&=&\frac{1}{3}\delta_{\alpha\beta}\\
&&-\frac{4}{9}\Big[(\hat n-1)\vec S\cdot\vec\sigma_{\alpha\beta}+i\eps^{abd}S^ds^c(\sigma^a\sigma^c\sigma^b)_{\alpha\beta}\Big]\nonumber
\eea
Now, we have
\be
\sigma^a\sigma^c=\delta^{ac}\bb 1+i\eps^{acf}\sigma^f
\ee
from which we conclude
\be
 \sigma^a\sigma^c\sigma^b=(\delta^{ac}\sigma^b+\delta^{bc}\sigma^a-\delta^{ab}\sigma^c)+i\eps^{acb}\bb 1.\label{eq3pauli}
\ee
When inserted into \eqref{acomm2}, the terms in the parenthesis give zero (last term vanishes under antisymmetrization and the first two cancel one-another), so that 
\bea\label{acomm3}
\{F\dn_\alpha,F\dg_\beta\}&=&\frac{1}{3}\delta_{\alpha\beta}\\
&&-\frac{4}{9}\Big[(\hat n-1)\vec S\cdot\vec\sigma_{\alpha\beta}-\eps^{abd}\eps^{acb}S^ds^c\delta_{\alpha\beta}\Big]\nonumber
\eea
Now since $\eps^{abd}\eps^{acb}=-2\delta^{cd}$ and $2\vec s=c\dg\vec\sigma c$ we have
\be
\{F\dn_\alpha,F\dg_\beta\}=\frac{1}{3}\delta_{\alpha\beta}-\frac{4}{9}(\hat n-1)\vec S\cdot\vec\sigma_{\alpha\beta}-\frac{4}{9}(\vec S\cdot c\dg\vec\sigma c)\delta_{\alpha\beta}
\ee
which we can write as
\bea
\{F\dn_\alpha,F\dg_\beta\}=\delta_{\alpha\beta}\\
&&\hspace{-.5cm}-\frac{4}{9}\Big[(\vec S\cdot c\dg\vec\sigma c+3/2)\delta_{\alpha\beta}+(\hat n-1)\vec S\cdot\vec\sigma_{\alpha\beta}\Big]\nonumber.
\eea
In the strong-coupling limit, the first and second term inside the square brackets vanish, and in this limit the anticommutator is normalized to unity.

\section{Equivalence between the F-spin and local moment at strong coupling. }
The spin operator of the F-electrons is
\be
\vec S_F=F\dg\frac{\vec\sigma}{2}F
\ee
Using \pref{eq:A1}, we can write the $c$ component of $\vec S_F$ as
\be
S_F^c=\frac{2}{9}c\dg_{\alpha'}c\dn_{\beta'}(\sigma^a\sigma^c\sigma^b)_{\alpha'\beta'}S^aS^b
\ee
Using Eq.\,\pref{eq3pauli} for the terms in parenthesis and Eq.\,\pref{eqS2} for $S^aS^b$ we find
\be
\vec S_F=\frac{2}{9}\Big[-c\dg\frac{\vec\sigma}{4}c+\vec S\Big]
\ee
At the strong Kondo coupling limit $(\vec S+c\dg\frac{\vec\sigma}{2}c)\ket{\Omega}=0$. Therefore, we conclude that as long as $\vec S_F$ acts on the product-state ground state
\be
\vec S_F=\frac{1}{3}\vec S.
\ee
\section{Appendix: Sum-rule on wavefunction weight}
\be
\Psi=\mat{\psi_c \\ \psi_F}, \quad {\cal G}(x,x';t)=\mat{G_{cc} & G_{cF} \\ G_{Fc} & G_{FF}}\nonumber
\ee
\be
\Psi(x,t)=\int{dx'}{\cal G}(x,t;x',0)\Psi(x',0)
\ee
Then, the total weight is
\bea
&&\int{dx}\psi\dg(x,t)\Psi(x,t)\\
&&=\int{dx'_1dx'_2}\Psi\dg(x'_1,0)\int{dx}{\cal G}\dg(x,t;x'_1){\cal G}\dn(x,t;x'_2)\Psi(x'_2,0)\nonumber
\eea
\section{Strong coupling expansion - Single-particle excitations}\label{sec:scsingle}
At strong coupling, the ground state is
\be
\ket{\phi}_0=\prod_j\ket{K_j}, \qquad \ket{K_j}=\frac{1}{\sqrt 2}\ket{\Ua_{j}\da_j-\Da_{j}\ua_j},
\ee
where $\Uparrow$ and $\Downarrow$ refer to the spins (magnetic moments) and $\ua$ and $\da$ refer to conduction electrons. To determine the energy of this state, note that
\bea
H_K&=&2J\sum_j\vec S_j\cdot\vec s_j\nonumber\\
&=&\sum_j\left\{\bmx{lll}J/2 \qquad\quad & S_j=1, & n_j=1 \\ 0 & S_j=1/2, & n_j\neq 1 \\ -3J/2 & S_j=0, & n_j=1.\emx\right.
\eea
Therefore the state $\ket{\phi}_0$ has energy $E_0/L=-3J_K/2$ where $L$ is the length of the system. The action of $H_t$ on $\ket{\phi_0}$ creates doublon-holon pairs $\ket{C_{n,n+1}}$ whose corresponding spins are in a spin-singlet, i.e.
\be
H_t\ket{\phi}_0=-t_c\sum_n \ket{S_{n+1,n}}\ket{C_{n+1,n}}\prod_{j\neq n,n+1}\ket{K_j},
\ee
where
\bea
\ket{C_{n+1,n}}=\frac{\ket{2_{n+1}0_n}+\ket{0_{n+1}2_n}}{\sqrt 2},\\ \ket{S_{n+1,n}}=\frac{\ket{\Ua_{n+1}\Da_n}-\ket{\Da_{n+1}\Ua_n}}{\sqrt{2}}
\eea
This excited  state has energy $E_\lambda=E_0+3J$, so  the second-order correction to the strong-coupling
ground state energy is
\be
\Delta E=-\sum_{\lambda\neq 0}\frac{\braket{\phi_0\vert H_t\vert\lambda}\braket{\lambda\vert H_t\vert \phi_0}}{E_\lambda-E_0}=-t^2\frac{1}{3J}\times N_s
\ee
leading to the energy $E_g/N=-{3J}/{2}-{t^2}/{3J}$. The correction to the wavefunction is
\bean
\ket{\phi}_1&=&\ket{\phi}_0+\sum_\lambda\frac{1}{E_0-E_\lambda}\ket{\lambda}\braket{\lambda\vert H_t\vert \phi}_0\\
&=&\Big[1+\frac{t}{3J}\sum_n\ket{S_{n,n+1};C_{n,n+1}}\bra{K_n;K_{n+1}}\Big]\ket{\phi}_0
\eean
i.e. there will be virtual doublon-holon pairs $\ket{C_{n+1,n}}$ whose corresponding spins are in a singlet state $\ket{S_{n+1,n}}$.
What are the single quasi-particle excitations of this ground state? We act on the ground state with
\be
c\dg_{n\sigma},\quad\text{and}\quad F_{n\sigma}\dg=\frac{2}{3}[\tilde\sigma S^z_n c\dg_{n,\sigma}+S_n^{\tilde{\sigma}
} c\dg_{n,-\sigma}]
\ee
Note that $\{c_{n\sigma},F\dg_{n\sigma}\}=\tilde{\sigma}S^z_n$,
where $\tilde\sigma=\pm$ for $\sigma=\ua,\da$. Assuming $k$ is a good quantum number we can 
find
\bea
{\sqrt 2}c\dg_{k\sigma}\ket{\phi}_1=(1-\frac{\eps_k}{6J})\ket{2;k,\sigma}_1\\
\sqrt 2 \tilde\sigma c\dn_{k\sigma}\ket{\phi}_1=(1-\frac{\eps_k}{6J})\ket{0;k,\sigma}_1\\
\sqrt{2}F\dg_{k\sigma}\ket{\phi}_1=(1+\frac{\eps_k}{6J})\ket{2;k,\sigma}_1\\  \sqrt{2}\tilde\sigma F_{k\sigma}\dg\ket{\phi}_1=(1+\frac{\eps_k}{6J})\ket{0;k,\sigma}_1
\eea
in terms of single doublon and holon states defined as
\bea
\ket{2;k,\sigma}=\sqrt{2}c\dg_{k\sigma}\ket{\phi_0}=\sum_m\varphi_k(m)\ket{2;m,\sigma}\\
\ket{0;k,\sigma}=\sqrt{2}c\dn_{k\sigma}\ket{\phi_0}=\sum_m\varphi_k(m)\ket{0;m,\sigma}
\eea
with energy
\be
\bra{2/0;k,\sigma}(H_0+H_t)\ket{2/0;k,\sigma}=E_0+\frac{3J}{2}+\frac{1}{2}\eps_k
\ee
\begin{figure}[tp]
\includegraphics[width=1\linewidth]{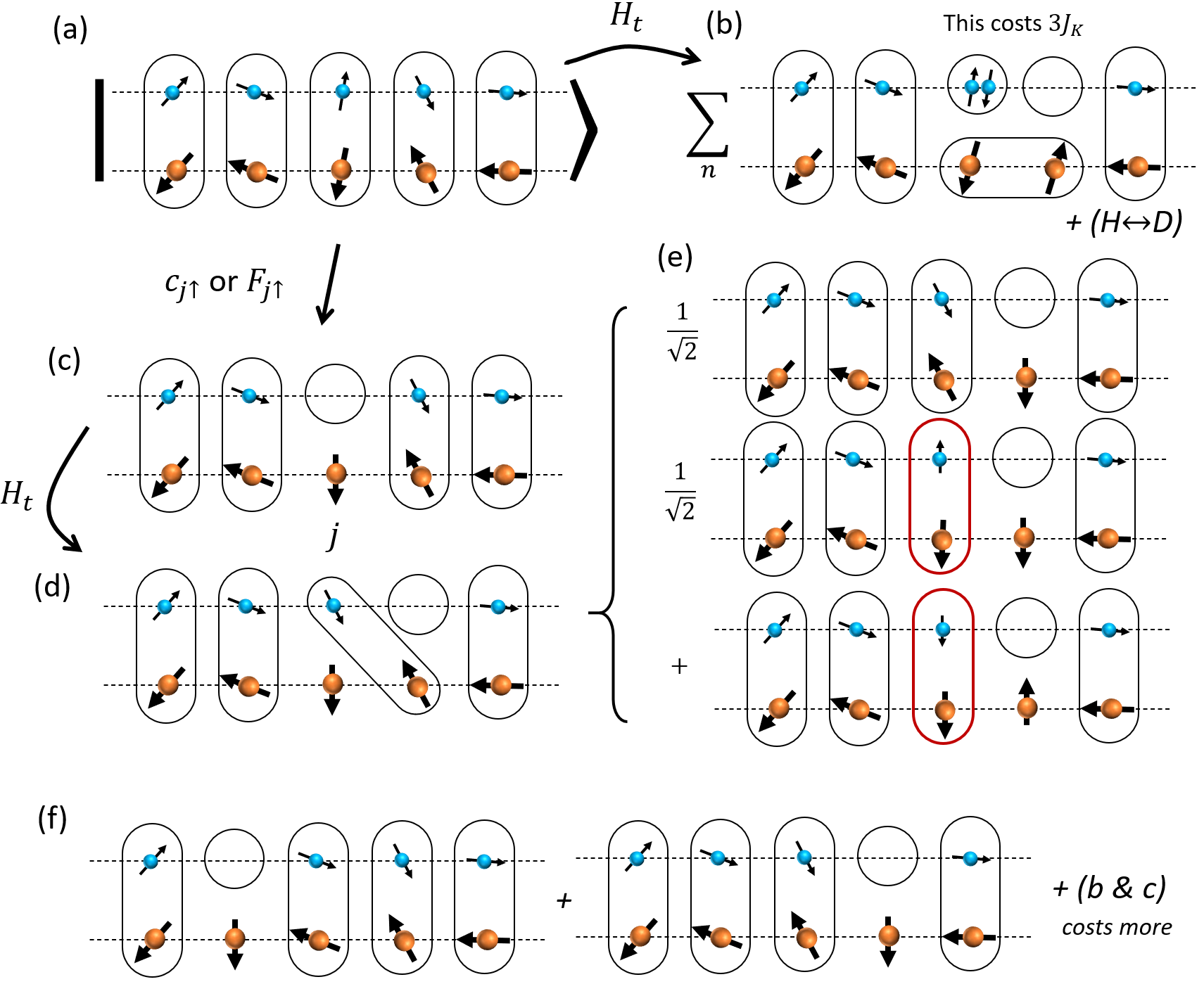}
\caption{\small Graphical illustration of what is done here. (a) $\ket{\psi_g}$ is the strong coupling ground state. (b) the result of acting with $H_t$ on $\ket{\psi_g}$. (c) Acting with $c_{j\sigma}$ or $F_{j\sigma}$ creates a holon. (d) The holon moves due to $H_t$ moving around singlets, but also high-energy triplets. (f) The final result after projection to low-energy singlet sector.}
\end{figure}
Using these and the spectral representation
\bea
G_{\alpha\beta}(z;k)&=&\frac{\sbraket{\phi\vert \psi_\alpha \vert 2;k,\sigma}\sbraket{2;k,\sigma\vert \psi\dg_\beta \vert \phi}}{z-[E_2(k)-E_\phi]}\nonumber\\
&&+\frac{\sbraket{0;k,\sigma\vert \psi_\alpha (k)\vert \phi}\sbraket{\phi\vert \psi\dg_\beta \vert 0;k,\sigma}}{z+[E_0(k)-E_\phi]}\qquad
\eea
we find the Green's function
\bean
\mat{G_{cc} & G_{cf} \\ G_{fc} & G_{ff}}_{(k,z)}&=&\frac{1}{z-(V+\eps_k/2)}\mat{1+\frac{\eps_k}{2V} & 1 \\ 1 & 1-\frac{\eps_k}{2V}}\nonumber\\
&&+\frac{1}{z+(V-\eps_k/2)}\mat{1-\frac{\eps_k}{V} & -1 \\ -1 & 1+\frac{\eps_k}{V}}
\eean
where $V\equiv 3J/2$. This can be written as 
\be
G(k,z)=\frac{1}{(z-\eps_k/2)^2-V^2}\mat{z & V \\ V & z-\eps_k}.
\ee
So to lowest order in $t$ we get
\be
G^{-1}(k,z)=z\bb 1-H, \qquad H=\mat{\eps_k & V \\ V & 0},
\ee
and we can interpret $V$ as a hybridization between the conduction and
composite f-electron. 
The real-space Green's function can be computed from $G(z,k)$ via
\be
G_{\alpha\beta}(z;x_1,x_2)=\sum_{k}\varphi_k(x_1)\varphi^*_{k}(x_2)G_{\alpha\beta}(z;k)
\ee
\section{Strong coupling expansion - Two-particle excitations\label{sec:twopsc}}
Single particle excitations are holons and doublons. The corresponding wavefunctions are
\be
\ket{D}_\sigma=\sum_n\psi_d(n)c\dg_{n\sigma}\ket{\Omega}, \quad \ket{H}_\sigma=\sum_n\psi_h(n)c\dn_{n\sigma}\ket{\Omega}\nonumber
\ee
and these have the energies $E_{d/h}(k)=E_1\pm\eps(k)$ where $\eps(k)=-t_c\cos(k)$ and $E_1=E_0+3J_K/2+U$.

Spin-excitations belong to the two-particle excitation spectrum. A $T^{+1}$ spin-triplet excitation has the wavefunction
\be
\ket{F^+}=\sum_{n_1n_2}\psi^+(n_1,n_2)c\dg_{n_1\ua}c\dn_{n_2\da}\ket{\Omega}
\ee
Such a state has energies around $E_2=E_0+3J_K+2U$. By acting on this with the Hamiltonian $H\ket{F^+}=E\ket{F^+}$ and projecting to stay within two-particle excitations, we find that the wavefunction $\psi(n_1,n_2)$ obeys the first-quantized Schr\"odinger equation
\bea
&&\frac{t}{2}\Big[\psi(n_1+1,n_2)+\psi(n_1-1,n_2)-\psi(n_1,n_2+1)\\
&&-\psi(n_1,n_2-1)]+U'\delta_{n_1,n_2}\psi(n_1,n_2)=(E-E_2)\psi(n_1,n_2)\nonumber
\eea
We solve this equation using the following ansatz
\bea
\psi_{k_1k_2}(n_1,n_2)&=&\theta(n_1<n_2)[Ae^{ik_1n_1-ik_2n_2}+A'e^{ik'_1n_1-ik'_2n_2}]\nonumber\\
&+&\theta(n_2<n_1)[Be^{ik_1n_1-ik_2n_2}+B'e^{ik'_1n_1-ik'_2n_2}]\nonumber\\
&+&\delta_{n_1n_2}Ce^{i(k_1-k_2)n_1}.
\eea
This wavefunction is labelled by quantum numbers $k_1$ and $k_2$ for doublon and holon respectively. However, note that in a typical scattering event $k_1$
$k'_i=\pi-k_i$. By plugging this wavefunction into the Schr\"odigner equation, we find that the doublon-holon pair state has energy
\be
E_{dh}({k_1,k_2})=E_2-t_c\cos(k_1)+t_c\cos(k_2)
\ee
Furthermore, $C=A+A'=B+B'$ and
\be
\mat{B \\ B'}=\mat{1+u & u \\ -u & 1-u}\mat{A\\ A'}, \label{equ}
\ee
where
\be
u_{k_1k_2}\equiv\frac{V/2it}{\sin k_2-\sin k_1}.
\ee
The right-hand side has to be this form, because after $L$ shift to the right $n_1+L>n_2$. So, we find
\be
Ae^{-ik_1L}=B, \andd A'e^{ik_2L}(-1)^L=B'
\ee
But we could also go left with the holon. It folows that
\be
Ae^{-ik_2L}=B, \andd A'e^{ik_1L}(-1)^L=B'
\ee
Combining these equations we see that
\be
e^{i(k_1-k_2)L}=1
\ee
Comparing this and the action of the translation operator on wavefunction, this is nothing but $(\hat P)^L=1$. Using these, the Schrodinger equation becomes
\be
M=\mat{1+u & u \\ -u & 1-u}-\mat{e^{-ik_1L}\\ & e^{ik_2L}(-1)^L},
\ee
$M\mat{A \\ A'}=0$. The $\det M=0$ gives
\be
2-(e^{-ik_1L}+e^{ik_2L})=-u_{k_1k_2}(e^{-ik_1L}-e^{ik_2L})
\ee
We can also find the corresponding eigenvector:
\be
A'=-\frac{1-e^{-ik_1L}}{1-e^{ik_2L}}A=e^{-ik_1L}A
\ee
where we have used $e^{i(k_1-k_2)L}=1$. This can be used to find that $A'=B$ and $B'=A$. 
We also have
\be
C=A+A'=A(1+e^{-ik_1L})
\ee
This fixes the wavefunction up to a normalization factor which is easily determined. So, we choose $A=1/\sqrt 2L$. In the following, we label the doublon-holon state with center of mass and relative momenta $\bar k$ and $p$ respectively.
\be
k_1=\bar k+p/2, \qquad k_2=-\bar k+p/2.
\ee
\subsubsection*{Spin-susceptibility}
We are interesting in computing the zero-temperature dynamic spin susceptibility defined by
\bea
\chi_S(n,\tau)&\equiv&\braket{-T_\tau S^+_n(\tau)S^-_0},\\
&=&-\braket{\Omega\vert c\dg_{n\ua}(\tau)c\dn_{n\da}(\tau)c\dg_{0\da}c\dn_{0\ua}\vert\Omega}\label{eqB16}, 
\eea
In the second line we have used that
\be
\Big(\vec S_n+c\dg_n\frac{\vec \sigma}{2}c_n\Big)\ket{\Omega}=0
\ee
By plugging-in
\be
\bb 1=\sum_{\bar k,p}\ket{F_{\bar k,p}}\bra{F_{\bar k,p}}
\ee
we find
\bea
\chi_S(n,\tau)&=&\sum_{\bar k,p>0}e^{-\tau \Delta E_{2p}(\bar k)}\bra{\Omega} c\dg_{n\ua}c\dn_{n\da}\ket{F_{\bar k,p}} \bra{F_{\bar k,p}}c\dg_{0\da}c\dn_{0\ua}\ket{\Omega}\nonumber\\
&=&\frac{1}{4}\sum_{\bar k,p>0}e^{-\tau \Delta E_{2p}(\bar k)}\psi\dn_{\bar k,p}(n,n)\psi^*_{\bar k,p}(0,0)\nonumber\\
&=&\frac{1}{4}\sum_{\bar k,p>0}e^{-\tau \Delta E_{2p}(\bar k)}e^{2i\bar k n}\abs{C_{\bar k,p}}^2\nonumber
\eea
where
\bea
\Delta E_{2p}(\bar k)&\equiv&E_{dh}(\bar k+p/2,-\bar k+p/2)-E_0\nonumber\\
&=&E_2-E_0+2t_c\sin(\bar k)\sin(p/2).
\eea
When taking Fourier transform 
\be
\chi_S(q,i\nu_p)=\sum_ne^{-iqn}\intob{d\tau e^{i\nu_p}}G_n(i\nu_p)
\ee
$2\bar k=k_1-k_2=q$ becomes (the positive freq. part only)
\be
\chi_S(q,i\nu_p)=\sum_{p>0}\frac{\abs{A_p(-q/2)}^2\cos^2(k_1L/2)}{i\nu_p-\Delta E_{2p}(-q/2)}.
\ee
\section{Mean-field theory}\label{sec:mft}
Representing the spin in Eq.\,(1) with fermionis $S_{\alpha\beta}=f\dg_\alpha f_\beta$ along with a constraint $f_\alpha\dg f_\alpha=1$ and decoupling the resulting four-fermion interaction using a Hubbard-Stratonovitch transformation, we arrive at
\be
\hspace{-.15cm}H=\sum_{k\sigma}\mat{c\dg_{k\sigma} & f\dg_{k\sigma}}\mat{\eps_c & V \\ V & \eps_f}\mat{c_{k\sigma} \\ f_{k\sigma}}+\frac{V^2}{J_K}+\lambda Q_f,\label{eqcki}
\ee
where $\eps_c=-2t_c\cos k$ and $\eps_f=0$ and the Lagrange multiplier $\lambda$ imposes the constraint on average. At p-h symmetry, considered here, $\lambda=0$.  The Hamiltonian\,\pref{eqcki} can be diaganalized using a SO(2) rotation
\be
\mat{c_{k\sigma}\\ f_{k\sigma}}=\mat{\cos\alpha_k & -\sin\alpha_k \\ \sin\alpha_k & \cos\alpha_k}\mat{l_{k\sigma} \\ h_{k\sigma}}
\ee
and the eigenenergies are
\be
E^{l/h}_k=\frac{\eps_k^c+\eps_k^f}{2}\pm\sqrt{\Big(\frac{\eps_k^c-\eps_k^f}{2}\Big)^2+V^2}.
\ee
Due to $\pi$-periodicity of the $\tan 2\alpha_k$, we are free to choose either the period $2\alpha_k\in (0,\pi)$ or $2\alpha_k\in (-\pi/2,\pi/2)$. We choose the former interval, because the angle evolves more continuously in the Brillion zone. Therefore,
\be
\sin2\alpha_k=\frac{2V}{E_k^l-E_k^h}, \qquad 
\cos2\alpha_k=\frac{\eps_k^c-\eps_k^f}{E_k^l-E_k^h},
\ee
The relation between Kondo couling and the dynamic hybridization is given by
\be
\frac{1}{J}=-\partial_{V^2}\sum_kE_k^h=\frac{1}{L}\sum_k\frac{1}{\sqrt{(\eps_k^c-\eps_k^f)^2+4V^2}}.
\ee
Assuming $\eps_f=0$, $\eps_c=-2t_c\cos k$, in the continuum limit,
\be
J/t_c={\pi}\frac{\sqrt{(V/t_c)^2+1}}{K(1/\sqrt{(V/t_c)^2+1})}.
\ee
where $K(k)$ is the complete elliptic integral of the first kind. The strong-coupling (large $V$) limit of this integral is $V\to J/2$. 
We can use this mean-field theory to compute the retarded Green's function
\be
G_f(k,\omega+i\eta)=\frac{\sin^2\alpha_k}{\omega+i\eta-E_l(k)}+\frac{\cos^2\alpha_k}{\omega+i\eta-E_h(k)}
\ee
as well as (anti-)time-ordered Green's functions
\bea
G_f^T(n,n';t>0)&=&\braket{-if(n,t)f\dg(n')}\nonumber\\
&=&\frac{-i}{L}\sum_k\phi_k(n)\phi_k^*(n')\sin^2\alpha_ke^{-iE^l_kt}\nonumber\\
G_f^{\tilde T}(n,n';t>0)&=&\braket{if\dg(n') f(n,t)}\nonumber\\
&=&\frac{i}{L}\sum_k\phi_k(n)\phi_k^*(n')\cos^2\alpha_ke^{-iE^h_kt}.\qquad
\eea
with 
\be
\phi_k(n)=\sqrt{\frac{2}{N}}\sin(nk).
\ee
These together with $G^R=\theta(t)(G^T-G^{\tilde T})$.\\
\section{Two-particle excitations - Random phase approximation}
In the non-interacting limit, the only contribution to Eq.~\pref{eqB16} 
is the disconnected part coming from Wick's contraction
\be
\chi_S(q,\tau)=\braket{-T_\tau c\dn_{n\da}(\tau)c\dg_{0\da}}\braket{T_\tau c\dg_{n\ua}(\tau)c_{0\ua}}
\ee
For non-interacting systems, we get the usual result
\be
\chi_S^0(q,\omega+i\eta)=\sum_k\frac{f(\eps_{k+q})-f(\eps_k)}{\omega+i\eta+\eps_{k+q}-\eps_k}
\ee
Therefore, we could assume that this is just the non-interacting $G_q^0(\tau)$ but multiplied by the factor $e^{-(3J_K+2U)\tau}$. Furthermore, the hopping of holons and doublons is exactly the same. So, we propose
\be
\chi_S^{dis.}(q,\omega+i\eta)=\sum_k\frac{1}{\omega+i\eta-(E^d_{k+q}+E^h_k)}
\ee
where 
\be
E^d_k=\frac{3J}{2}+U-t_c\cos k, \qquad E^h_k=\frac{3J}{2}+U+t_c\cos k
\ee
This is shown in the figure. However, the magnon band is missing, even if we include RPA:
\be
\chi_S^{\rm RPA}(q,\omega+i\eta)=\frac{1}{[\chi_S^{\rm dis}(q,\omega+i\eta)]^{-1}-U'}
\ee

\section{MPS Methods for Computing Green Functions \label{MPSGF} }
In order to calculate Green's function $G(q,\omega)$, spin susceptibility $\chi(q,\omega)$ and $F(q,\omega)$, 
we first calculate the ground state by the DMRG method.   The ground state is represented by $2N$ site matrix product state (MPS).  The spin sites of dimension 2 are located on the odd sites with the remaining conduction electrons  on the even sites of dimension 4.  The bond dimensions are determined automatically, and controlled by a truncation error threshold or cutoff of $\epsilon = 10^{-12}$ in the ground state calculation. 

The Green's function are defined in \eq{eq:green}. To obtain the retarded $G^R(x,t)$ and  $\chi(x,t)$,  we need to calculate different components of Green's function, such as $G^>_{cc}$, $G^>_{cF}$,  $G^<_{cc}$, etc. These terms come out because there are both  $F$ and $c$ degrees of freedom, and also we need to both greater and lesser Green's function to compute the retarded Green's function. Without loss of generality,  we take $G^>_{cF}$  as an example. The other components are calculated by a similar approach. 
\begin{eqnarray}
G^>_{ cF}  &=& -i \braket{ c_{x_1}(t)F_{x_2}^\dag (0) } \\
&=& -i  \braket{ c_{x_1}(t/2)F_{x_2}^\dag (-t/2) } \\
&=& -i \braket{ e^{{iHt}/{2}} c_{x_1} e^{{-iHt}/{2}} e^{{-iHt}/{2} }F_{x_2}^\dag e^{{iHt}/{2}  }} \\
&=& -i e^{iE_0 t} \bra{0} c_{x_1} e^{{-iHt}/{2}}   e^{{-iHt}/{2} }F_{x_2}^\dag\ket{0 }
\end{eqnarray}
where $E_0$ is the ground state energy.  The second equal sign holds because translational invariance of $t$. 
From the third line, we choose the Heisenberg picture. It can written into the overlap of two time evolving MPS.
\begin{eqnarray}
G^>_{ cF}  &=&  -i e^{iE_0 t} \left( e^{{iHt}/{2}} c_{x_1}^\dag\ket{ 0 } \right)^\dag \left( e^{{-iHt}/{2} }F_{x_2}^\dag  \ket{ 0} \right)
\end{eqnarray}

We first apply one on-site operator $c_{x_1}^\dag$ ( $F_{x_2}^\dag )$ to the ground state.  Two groups of MPS are obtained depending on $x_1$ and $x_2$,  then we use the time evolving block decimation (TEBD) \cite{Vidal04,White04} to time evolve for $+\frac{t}{2}$ and $-\frac{t}{2}$. In our calculation, instead of time evolving the right ket for $t$, we evolve the bra for $-t/2$ and ket for $t/2$. Recalling that the bond dimension of an MPS generically grows exponentially under real-time evolution, splitting the time and equally distributing gates onto the bra and ket allows us to work with two MPS with significantly smaller bond dimension rather than one MPS with a large bond dimension.  
We obtain $u(x,\frac{t}{2})$ and $v(x,\frac{t}{2})$, which are defined as
\begin{eqnarray}
\ket{u(x,t/2)} &=&  e^{{iHt}/{2}} c_{x_1}^\dag \ket{ 0 } \\
\ket{v(x,t/2)} &=& e^{{-iHt}/{2} }F_{x_2}^\dag \ket{ 0 } .
\end{eqnarray}
The total time slot $t/2$ is split into many time slice of $\tau$.  For every time step, we compute the Green function by calculate the overlap between MPS. 

The time evolution reaches a certain time $T_{max}$. The resolution in frequency domains depends on $T_{max}$. $\Delta \omega = \frac{1}{T_{max}}$. The longer time we run, we get more find details in $\omega$. In our calculation, we checked the measurement results from different $T_{max}$ and  confirmed our results  converged when $T_{max} > 100$.    

The MPS bond dimension grows so the MPS bond dimension requires truncation. During the time evolving, we have tried different cut off error from $\epsilon=10^{-4}$ to $10^{-8}$.  As a result, the green function for $\epsilon=10^{-5}$ has already converged.  

Having obtained $G^>$ and $G^<$, the retarded green function can be derived by
\begin{equation}
 G^R(x_1,x_2,t) = \Theta(t) \big( G^>(x_1,x_2,t) - G^<(x_1,x_2,t) \big) \ .
 \end{equation}
We can calculate the retarded Green's function $G^R(q,\omega)$ in the $(q,\omega)$ domain by Fourier transform 
\begin{equation}
 G^R(q,\omega) = \int dt\sum_{x_1,x_2} e^{i q (x_1 - x_2 ) - i \omega t } G^R(x_1,x_2,t). \label{eq:fourier} 
 \end{equation}

\section{Spin-susceptibility within mean-field theory}
We can write the spin-susceptibility as
\bea
\chi_{ff}^{+-}(n,\tau)&=&\braket{-T_\tau S_n^+(\tau)S_0^-(0)}\\
&=&\braket{-T_\tau f\dg_{n\ua}(\tau)f\dn_{n\da}(\tau)f\dg_{0\da}f\dn_{0\ua}}\nonumber\\
&=&\braket{-T_\tau f_{n\da}(\tau)f_{0\da}\dg}\braket{T_\tau f\dg_{n\ua}(\tau)f\dn_{0\ua}},\nonumber
\eea
so that
\be
\chi_{ff}^{+-}(q,\tau)=\sum_kG_f(k+q,\tau)G_f(k,-\tau).
\ee
Going to (retarded real) frequency domain
\bea
\chi(q,\omega+i\eta)&=&-\sum_k\int{\frac{dx}{\pi}}f(x)G''(k,x)\times\\
&&\hspace{-1.5cm}\Big\{G_f(x+\omega+i\eta,k+q)\nonumber+G_f(x-\omega-i\eta,k-q)\Big\}\nonumber.
\eea
For the mean-field calculations, we use the notation in Appendix B of \cite{Wugalter2020}. 
The $G_f$ within mean-field is given by
\be
G_f(k,\omega+i\eta)=\frac{\sin^2\alpha_k}{\omega+i\eta-E_+(k)}+\frac{\cos^2\alpha_k}{\omega+i\eta-E_-(k)}
\ee
where $\alpha_k$, $E_\pm(k)$ are defined in the appendix. Plugging-in Eq.\,(1) we find
\bea
\chi^{+-}_{ff}(q,\omega+i\eta)&=&\sum_k\cos^2\alpha_k\times\\
&&\hspace{-0.5cm}\Big[\frac{\sin^2\alpha_{k+q}}{\omega+i\eta+E^-(k)-E^+(k+q)}\nonumber\\
&&\hspace{0.5cm}-\frac{\sin^2\alpha_{k-q}}{\omega+i\eta-E^-(k)+E^+(k+q)}\Big]\nonumber
\eea
Both term are there, but for $\omega>0$ only the first term contributes to the imaginary part
\bea
\chi''_{ff}(q,\omega>0)&=&\pi\sum_k\cos^2\alpha_k\sin^2\alpha_{k+q}\times\\
&&\qquad\delta\Big({\omega+E_-(k)-E_+(k-q)}\Big)\nonumber
\eea
There were some numerical errors previously. Next, we plot this function assuming $\eps_f=0$.

\begin{figure}[h!]
\includegraphics[width=1\linewidth]{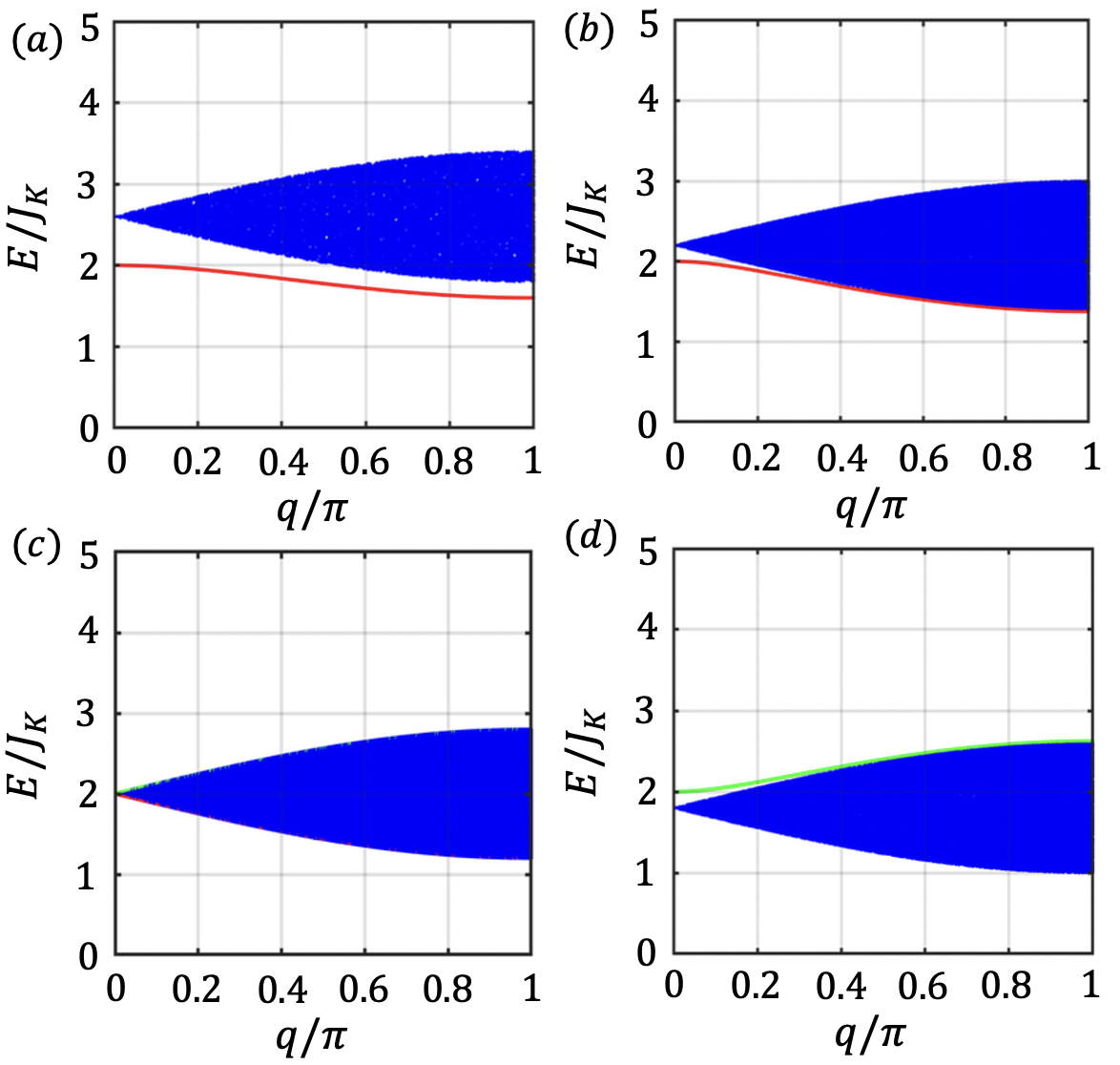}
\caption{\small Dynamical spin susceptibility from the strong coupling expansion. (a) $U=0$ (b) $U=..$ (c) $U=..$ (d) $U=..$}

\end{figure}
We can also do RPA on this. We define
\be
\chi_{RPA}(q,\omega)=\frac{1}{\chi^{-1}(q,\omega)-W}
\ee

\section{Parallelization of MPS Calculations}
The Green's function at a given time is a matrix defined in $(x_1,x_2)$ domain. The calculation of each entry $(x_1,x_2)$ involves independent time-evolution calculations and overlaps of different wave functions $\left|u\left(x,t\right)\right\rangle$s and $\left|u\left(x,t\right)\right\rangle$s. These wavefunctions originate with creation and annihilation operators acting on different sites $x)$. So we can parallelize these calculations and significantly reduce the time to solution. For each time slice or value of $t$, the computation contains two parts, the time evolution and measurement.

The time evolution of $\left|u\left(x,t\right)\right\rangle$s and $\left|u\left(x,t\right)\right\rangle$s are independent of each other and consume approximately same amount of time, which can run in different threads with minor data exchange. In total, there are $O(N)$ number of wave functions, which can be parallelized with no overhead cost, and scale well with increasing number of threads.

The measurement of the Green's functions matrix involves calculating the overlap of $\left|u\left(x,t\right)\right\rangle$s at different sites $x_1$ and $x_2$. Both $x_1$ and $x_2$ run from 1 to $N$. And the computation of these overlaps are independent, which can be computed with $O(N^2)$ threads. 

The time evolution step takes the dominant amount of time, because each time evolution requires application of series of gates and repeating singular value decomposition to keep bond dimension increasing, which contributes to a large prefactor before $O(N)$. Though the measurement scales as $O(N^2)$, the overlap operation is much faster.

\bibliography{main}

\providecommand{\noopsort}[1]{}\providecommand{\singleletter}[1]{#1}%
\begin{thebibliography}{47}%
\makeatletter
\providecommand \@ifxundefined [1]{%
 \@ifx{#1\undefined}
}%
\providecommand \@ifnum [1]{%
 \ifnum #1\expandafter \@firstoftwo
 \else \expandafter \@secondoftwo
 \fi
}%
\providecommand \@ifx [1]{%
 \ifx #1\expandafter \@firstoftwo
 \else \expandafter \@secondoftwo
 \fi
}%
\providecommand \natexlab [1]{#1}%
\providecommand \enquote  [1]{``#1''}%
\providecommand \bibnamefont  [1]{#1}%
\providecommand \bibfnamefont [1]{#1}%
\providecommand \citenamefont [1]{#1}%
\providecommand \href@noop [0]{\@secondoftwo}%
\providecommand \href [0]{\begingroup \@sanitize@url \@href}%
\providecommand \@href[1]{\@@startlink{#1}\@@href}%
\providecommand \@@href[1]{\endgroup#1\@@endlink}%
\providecommand \@sanitize@url [0]{\catcode `\\12\catcode `\$12\catcode
  `\&12\catcode `\#12\catcode `\^12\catcode `\_12\catcode `\%12\relax}%
\providecommand \@@startlink[1]{}%
\providecommand \@@endlink[0]{}%
\providecommand \url  [0]{\begingroup\@sanitize@url \@url }%
\providecommand \@url [1]{\endgroup\@href {#1}{\urlprefix }}%
\providecommand \urlprefix  [0]{URL }%
\providecommand \Eprint [0]{\href }%
\providecommand \doibase [0]{https://doi.org/}%
\providecommand \selectlanguage [0]{\@gobble}%
\providecommand \bibinfo  [0]{\@secondoftwo}%
\providecommand \bibfield  [0]{\@secondoftwo}%
\providecommand \translation [1]{[#1]}%
\providecommand \BibitemOpen [0]{}%
\providecommand \bibitemStop [0]{}%
\providecommand \bibitemNoStop [0]{.\EOS\space}%
\providecommand \EOS [0]{\spacefactor3000\relax}%
\providecommand \BibitemShut  [1]{\csname bibitem#1\endcsname}%
\let\auto@bib@innerbib\@empty
\bibitem [{\citenamefont {Menth}\ \emph {et~al.}(1969)\citenamefont {Menth},
  \citenamefont {Buehler},\ and\ \citenamefont {Geballe}}]{Menth_PRL1969}%
  \BibitemOpen
  \bibfield  {author} {\bibinfo {author} {\bibfnamefont {A.}~\bibnamefont
  {Menth}}, \bibinfo {author} {\bibfnamefont {E.}~\bibnamefont {Buehler}},\
  and\ \bibinfo {author} {\bibfnamefont {T.~H.}\ \bibnamefont {Geballe}},\
  }\bibfield  {title} {\bibinfo {title} {Magnetic and semiconducting properties
  of $smb_6$},\ }\href@noop {} {\bibfield  {journal} {\bibinfo  {journal}
  {Phys. Rev. Lett.}\ }\textbf {\bibinfo {volume} {22}},\ \bibinfo {pages}
  {295} (\bibinfo {year} {1969})}\BibitemShut {NoStop}%
\bibitem [{\citenamefont {Mott}(1974)}]{Mott74}%
  \BibitemOpen
  \bibfield  {author} {\bibinfo {author} {\bibfnamefont {N.~F.}\ \bibnamefont
  {Mott}},\ }\bibfield  {title} {\bibinfo {title} {Rare-earth compounds with
  mixed valencies},\ }\href {https://doi.org/10.1080/14786439808206566}
  {\bibfield  {journal} {\bibinfo  {journal} {The Phil. Mag.}\ }\textbf
  {\bibinfo {volume} {30}},\ \bibinfo {pages} {403} (\bibinfo {year}
  {1974})}\BibitemShut {NoStop}%
\bibitem [{\citenamefont {Doniach}(1977)}]{Doniach77}%
  \BibitemOpen
  \bibfield  {author} {\bibinfo {author} {\bibfnamefont {S.}~\bibnamefont
  {Doniach}},\ }\bibfield  {title} {\bibinfo {title} {The kondo lattice and
  weak antiferromagnetism},\ }\href
  {https://doi.org/10.1016/0378-4363(77)90190-5} {\bibfield  {journal}
  {\bibinfo  {journal} {Physica B+C}\ }\textbf {\bibinfo {volume} {91}},\
  \bibinfo {pages} {231} (\bibinfo {year} {1977})}\BibitemShut {NoStop}%
\bibitem [{\citenamefont {Aeppli}\ and\ \citenamefont {Fisk}(1992)}]{Fisk92}%
  \BibitemOpen
  \bibfield  {author} {\bibinfo {author} {\bibfnamefont {G.}~\bibnamefont
  {Aeppli}}\ and\ \bibinfo {author} {\bibfnamefont {Z.}~\bibnamefont {Fisk}},\
  }\bibfield  {title} {\bibinfo {title} {{Kondo Insulators}},\ }\href
  {www.scopus.com} {\bibfield  {journal} {\bibinfo  {journal}
  {Commun.Condens.Matter}\ }\textbf {\bibinfo {volume} {16}},\ \bibinfo {pages}
  {155} (\bibinfo {year} {1992})}\BibitemShut {NoStop}%
\bibitem [{\citenamefont {Riseborough}(2000)}]{Riseborough2010}%
  \BibitemOpen
  \bibfield  {author} {\bibinfo {author} {\bibfnamefont {P.}~\bibnamefont
  {Riseborough}},\ }\bibfield  {title} {\bibinfo {title} {Heavy fermion
  semiconductors},\ }\href@noop {} {\bibfield  {journal} {\bibinfo  {journal}
  {Advances in Physics}\ }\textbf {\bibinfo {volume} {49}} (\bibinfo {year}
  {2000})}\BibitemShut {NoStop}%
\bibitem [{\citenamefont {Kasuya}(1956)}]{kasuya1955}%
  \BibitemOpen
  \bibfield  {author} {\bibinfo {author} {\bibfnamefont {T.}~\bibnamefont
  {Kasuya}},\ }\bibfield  {title} {\bibinfo {title} {{A Theory of Metallic
  Ferro- and Antiferromagnetism on Zener's Model}},\ }\href
  {https://doi.org/10.1143/PTP.16.45} {\bibfield  {journal} {\bibinfo
  {journal} {Progress of Theoretical Physics}\ }\textbf {\bibinfo {volume}
  {16}},\ \bibinfo {pages} {45} (\bibinfo {year} {1956})}\BibitemShut {NoStop}%
\bibitem [{\citenamefont {Lacroix}\ and\ \citenamefont
  {Cyrot}(1979)}]{lacroix79}%
  \BibitemOpen
  \bibfield  {author} {\bibinfo {author} {\bibfnamefont {C.}~\bibnamefont
  {Lacroix}}\ and\ \bibinfo {author} {\bibfnamefont {M.}~\bibnamefont
  {Cyrot}},\ }\bibfield  {title} {\bibinfo {title} {Phase diagram of the kondo
  lattice},\ }\href {https://doi.org/10.1103/PhysRevB.20.1969} {\bibfield
  {journal} {\bibinfo  {journal} {Phys. Rev. B}\ }\textbf {\bibinfo {volume}
  {20}},\ \bibinfo {pages} {1969} (\bibinfo {year} {1979})}\BibitemShut
  {NoStop}%
\bibitem [{\citenamefont {Coleman}(1984)}]{Coleman84}%
  \BibitemOpen
  \bibfield  {author} {\bibinfo {author} {\bibfnamefont {P.}~\bibnamefont
  {Coleman}},\ }\bibfield  {title} {\bibinfo {title} {New approach to the
  mixed-valence problem},\ }\href {https://doi.org/10.1103/PhysRevB.29.3035}
  {\bibfield  {journal} {\bibinfo  {journal} {Phys. Rev. B}\ }\textbf {\bibinfo
  {volume} {29}},\ \bibinfo {pages} {3035} (\bibinfo {year}
  {1984})}\BibitemShut {NoStop}%
\bibitem [{\citenamefont {Read}\ \emph {et~al.}(1984)\citenamefont {Read},
  \citenamefont {Newns},\ and\ \citenamefont {Doniach}}]{Read84}%
  \BibitemOpen
  \bibfield  {author} {\bibinfo {author} {\bibfnamefont {N.}~\bibnamefont
  {Read}}, \bibinfo {author} {\bibfnamefont {D.~M.}\ \bibnamefont {Newns}},\
  and\ \bibinfo {author} {\bibfnamefont {S.}~\bibnamefont {Doniach}},\
  }\bibfield  {title} {\bibinfo {title} {Stability of the kondo lattice in the
  large-$n$ limit},\ }\href {https://doi.org/10.1103/PhysRevB.30.3841}
  {\bibfield  {journal} {\bibinfo  {journal} {Phys. Rev. B}\ }\textbf {\bibinfo
  {volume} {30}},\ \bibinfo {pages} {3841} (\bibinfo {year}
  {1984})}\BibitemShut {NoStop}%
\bibitem [{\citenamefont {Auerbach}\ and\ \citenamefont
  {Levin}(1986)}]{Auerbach86}%
  \BibitemOpen
  \bibfield  {author} {\bibinfo {author} {\bibfnamefont {A.}~\bibnamefont
  {Auerbach}}\ and\ \bibinfo {author} {\bibfnamefont {K.}~\bibnamefont
  {Levin}},\ }\bibfield  {title} {\bibinfo {title} {Kondo bosons and the kondo
  lattice: Microscopic basis for the heavy fermi liquid},\ }\href
  {https://doi.org/10.1103/PhysRevLett.57.877} {\bibfield  {journal} {\bibinfo
  {journal} {Phys. Rev. Lett.}\ }\textbf {\bibinfo {volume} {57}},\ \bibinfo
  {pages} {877} (\bibinfo {year} {1986})}\BibitemShut {NoStop}%
\bibitem [{\citenamefont {Millis}\ and\ \citenamefont {Lee}(1987)}]{Millis87}%
  \BibitemOpen
  \bibfield  {author} {\bibinfo {author} {\bibfnamefont {A.~J.}\ \bibnamefont
  {Millis}}\ and\ \bibinfo {author} {\bibfnamefont {P.~A.}\ \bibnamefont
  {Lee}},\ }\bibfield  {title} {\bibinfo {title} {Large-orbital-degeneracy
  expansion for the lattice anderson model},\ }\href
  {https://doi.org/10.1103/PhysRevB.35.3394} {\bibfield  {journal} {\bibinfo
  {journal} {Phys. Rev. B}\ }\textbf {\bibinfo {volume} {35}},\ \bibinfo
  {pages} {3394} (\bibinfo {year} {1987})}\BibitemShut {NoStop}%
\bibitem [{\citenamefont {Coleman}()}]{Coleman87b}%
  \BibitemOpen
  \bibfield  {author} {\bibinfo {author} {\bibfnamefont {P.}~\bibnamefont
  {Coleman}},\ }\bibfield  {title} {\bibinfo {title} {{Constrained
  quasiparticles and conduction in heavy-fermion systems}},\ }\href
  {https://journals.aps.org/prl/abstract/10.1103/PhysRevLett.59.1026}
  {\bibfield  {journal} {\bibinfo  {journal} {Phys. Rev. Lett.}\ }\textbf
  {\bibinfo {volume} {59}},\ \bibinfo {pages} {1026}}\BibitemShut {NoStop}%
\bibitem [{\citenamefont {Danu}\ \emph {et~al.}(2021)\citenamefont {Danu},
  \citenamefont {Liu}, \citenamefont {Assaad},\ and\ \citenamefont
  {Raczkowski}}]{Danu2021}%
  \BibitemOpen
  \bibfield  {author} {\bibinfo {author} {\bibfnamefont {B.}~\bibnamefont
  {Danu}}, \bibinfo {author} {\bibfnamefont {Z.}~\bibnamefont {Liu}}, \bibinfo
  {author} {\bibfnamefont {F.~F.}\ \bibnamefont {Assaad}},\ and\ \bibinfo
  {author} {\bibfnamefont {M.}~\bibnamefont {Raczkowski}},\ }\bibfield  {title}
  {\bibinfo {title} {Zooming in on heavy fermions in kondo lattice models},\
  }\href {https://doi.org/10.1103/PhysRevB.104.155128} {\bibfield  {journal}
  {\bibinfo  {journal} {Phys. Rev. B}\ }\textbf {\bibinfo {volume} {104}},\
  \bibinfo {pages} {155128} (\bibinfo {year} {2021})}\BibitemShut {NoStop}%
\bibitem [{\citenamefont {Tsunetsugu}\ \emph {et~al.}()\citenamefont
  {Tsunetsugu}, \citenamefont {Sigrist},\ and\ \citenamefont
  {Ueda}}]{Tsunetsugu}%
  \BibitemOpen
  \bibfield  {author} {\bibinfo {author} {\bibfnamefont {H.}~\bibnamefont
  {Tsunetsugu}}, \bibinfo {author} {\bibfnamefont {M.}~\bibnamefont
  {Sigrist}},\ and\ \bibinfo {author} {\bibfnamefont {K.}~\bibnamefont
  {Ueda}},\ }\bibfield  {title} {\bibinfo {title} {{The ground-state phase
  diagram of the one-dimensional Kondo lattice model}},\ }\bibfield  {journal}
  {\bibinfo  {journal} {Reviews of Modern Physics}\ }\href
  {https://doi.org/10.1103/revmodphys.69.809}
  {10.1103/revmodphys.69.809}\BibitemShut {NoStop}%
\bibitem [{\citenamefont {Fye}\ and\ \citenamefont {Scalapino}(1990)}]{Fye90}%
  \BibitemOpen
  \bibfield  {author} {\bibinfo {author} {\bibfnamefont {R.~M.}\ \bibnamefont
  {Fye}}\ and\ \bibinfo {author} {\bibfnamefont {D.~J.}\ \bibnamefont
  {Scalapino}},\ }\bibfield  {title} {\bibinfo {title} {One-dimensional
  symmetric kondo lattice: A quantum monte carlo study},\ }\href
  {https://doi.org/10.1103/PhysRevLett.65.3177} {\bibfield  {journal} {\bibinfo
   {journal} {Phys. Rev. Lett.}\ }\textbf {\bibinfo {volume} {65}},\ \bibinfo
  {pages} {3177} (\bibinfo {year} {1990})}\BibitemShut {NoStop}%
\bibitem [{\citenamefont {Troyer}\ and\ \citenamefont
  {W\"urtz}(1993)}]{Troyer93}%
  \BibitemOpen
  \bibfield  {author} {\bibinfo {author} {\bibfnamefont {M.}~\bibnamefont
  {Troyer}}\ and\ \bibinfo {author} {\bibfnamefont {D.}~\bibnamefont
  {W\"urtz}},\ }\bibfield  {title} {\bibinfo {title} {Ferromagnetism of the
  one-dimensional kondo-lattice model: A quantum monte carlo study},\ }\href
  {https://doi.org/10.1103/PhysRevB.47.2886} {\bibfield  {journal} {\bibinfo
  {journal} {Phys. Rev. B}\ }\textbf {\bibinfo {volume} {47}},\ \bibinfo
  {pages} {2886} (\bibinfo {year} {1993})}\BibitemShut {NoStop}%
\bibitem [{\citenamefont {Raczkowski}\ and\ \citenamefont
  {Assaad}(2019)}]{Raczkowski19}%
  \BibitemOpen
  \bibfield  {author} {\bibinfo {author} {\bibfnamefont {M.}~\bibnamefont
  {Raczkowski}}\ and\ \bibinfo {author} {\bibfnamefont {F.~F.}\ \bibnamefont
  {Assaad}},\ }\bibfield  {title} {\bibinfo {title} {Emergent coherent lattice
  behavior in kondo nanosystems},\ }\href
  {https://doi.org/10.1103/PhysRevLett.122.097203} {\bibfield  {journal}
  {\bibinfo  {journal} {Phys. Rev. Lett.}\ }\textbf {\bibinfo {volume} {122}},\
  \bibinfo {pages} {097203} (\bibinfo {year} {2019})}\BibitemShut {NoStop}%
\bibitem [{\citenamefont {Yu}\ and\ \citenamefont {White}(1993)}]{Yu93}%
  \BibitemOpen
  \bibfield  {author} {\bibinfo {author} {\bibfnamefont {C.~C.}\ \bibnamefont
  {Yu}}\ and\ \bibinfo {author} {\bibfnamefont {S.~R.}\ \bibnamefont {White}},\
  }\bibfield  {title} {\bibinfo {title} {Numerical renormalization group study
  of the one-dimensional kondo insulator},\ }\href
  {https://doi.org/10.1103/PhysRevLett.71.3866} {\bibfield  {journal} {\bibinfo
   {journal} {Phys. Rev. Lett.}\ }\textbf {\bibinfo {volume} {71}},\ \bibinfo
  {pages} {3866} (\bibinfo {year} {1993})}\BibitemShut {NoStop}%
\bibitem [{\citenamefont {Sikkema}\ \emph {et~al.}(1997)\citenamefont
  {Sikkema}, \citenamefont {Affleck},\ and\ \citenamefont {White}}]{Sikkema97}%
  \BibitemOpen
  \bibfield  {author} {\bibinfo {author} {\bibfnamefont {A.~E.}\ \bibnamefont
  {Sikkema}}, \bibinfo {author} {\bibfnamefont {I.}~\bibnamefont {Affleck}},\
  and\ \bibinfo {author} {\bibfnamefont {S.~R.}\ \bibnamefont {White}},\
  }\bibfield  {title} {\bibinfo {title} {Spin gap in a doped kondo chain},\
  }\href {https://doi.org/10.1103/PhysRevLett.79.929} {\bibfield  {journal}
  {\bibinfo  {journal} {Phys. Rev. Lett.}\ }\textbf {\bibinfo {volume} {79}},\
  \bibinfo {pages} {929} (\bibinfo {year} {1997})}\BibitemShut {NoStop}%
\bibitem [{\citenamefont {McCulloch}\ \emph {et~al.}(1999)\citenamefont
  {McCulloch}, \citenamefont {Gulacsi}, \citenamefont {Caprara}, \citenamefont
  {Jazavaou},\ and\ \citenamefont {Rosengren}}]{McCulloch99}%
  \BibitemOpen
  \bibfield  {author} {\bibinfo {author} {\bibfnamefont {I.~P.}\ \bibnamefont
  {McCulloch}}, \bibinfo {author} {\bibfnamefont {M.}~\bibnamefont {Gulacsi}},
  \bibinfo {author} {\bibfnamefont {S.}~\bibnamefont {Caprara}}, \bibinfo
  {author} {\bibfnamefont {A.}~\bibnamefont {Jazavaou}},\ and\ \bibinfo
  {author} {\bibfnamefont {A.}~\bibnamefont {Rosengren}},\ }\bibfield  {title}
  {\bibinfo {title} {Phase diagram of the 1d kondo lattice model},\ }\href
  {https://doi.org/10.1023/a:1022557314114} {\bibfield  {journal} {\bibinfo
  {journal} {Journal of Low Temperature Physics}\ }\textbf {\bibinfo {volume}
  {117}},\ \bibinfo {pages} {323} (\bibinfo {year} {1999})}\BibitemShut
  {NoStop}%
\bibitem [{\citenamefont {Shibata}\ and\ \citenamefont
  {Ueda}(1999)}]{Shibata99}%
  \BibitemOpen
  \bibfield  {author} {\bibinfo {author} {\bibfnamefont {N.}~\bibnamefont
  {Shibata}}\ and\ \bibinfo {author} {\bibfnamefont {K.}~\bibnamefont {Ueda}},\
  }\bibfield  {title} {\bibinfo {title} {The one-dimensional kondo lattice
  model studied by the density matrix renormalization group method},\ }\href
  {https://doi.org/10.1088/0953-8984/11/2/002} {\bibfield  {journal} {\bibinfo
  {journal} {Journal of Physics: Condensed Matter}\ }\textbf {\bibinfo {volume}
  {11}},\ \bibinfo {pages} {R1} (\bibinfo {year} {1999})}\BibitemShut {NoStop}%
\bibitem [{\citenamefont {Peters}\ and\ \citenamefont
  {Kawakami}(2012)}]{Peters12}%
  \BibitemOpen
  \bibfield  {author} {\bibinfo {author} {\bibfnamefont {R.}~\bibnamefont
  {Peters}}\ and\ \bibinfo {author} {\bibfnamefont {N.}~\bibnamefont
  {Kawakami}},\ }\bibfield  {title} {\bibinfo {title} {Ferromagnetic state in
  the one-dimensional kondo lattice model},\ }\href
  {https://doi.org/10.1103/PhysRevB.86.165107} {\bibfield  {journal} {\bibinfo
  {journal} {Phys. Rev. B}\ }\textbf {\bibinfo {volume} {86}},\ \bibinfo
  {pages} {165107} (\bibinfo {year} {2012})}\BibitemShut {NoStop}%
\bibitem [{\citenamefont {Zachar}\ \emph {et~al.}(1996)\citenamefont {Zachar},
  \citenamefont {Kivelson},\ and\ \citenamefont {Emery}}]{Zachar96}%
  \BibitemOpen
  \bibfield  {author} {\bibinfo {author} {\bibfnamefont {O.}~\bibnamefont
  {Zachar}}, \bibinfo {author} {\bibfnamefont {S.~A.}\ \bibnamefont
  {Kivelson}},\ and\ \bibinfo {author} {\bibfnamefont {V.~J.}\ \bibnamefont
  {Emery}},\ }\bibfield  {title} {\bibinfo {title} {Exact results for a 1d
  kondo lattice from bosonization},\ }\href
  {https://doi.org/10.1103/PhysRevLett.77.1342} {\bibfield  {journal} {\bibinfo
   {journal} {Phys. Rev. Lett.}\ }\textbf {\bibinfo {volume} {77}},\ \bibinfo
  {pages} {1342} (\bibinfo {year} {1996})}\BibitemShut {NoStop}%
\bibitem [{\citenamefont {Tsvelik}\ and\ \citenamefont
  {Yevtushenko}(2019)}]{Tsvelik19}%
  \BibitemOpen
  \bibfield  {author} {\bibinfo {author} {\bibfnamefont {A.~M.}\ \bibnamefont
  {Tsvelik}}\ and\ \bibinfo {author} {\bibfnamefont {O.~M.}\ \bibnamefont
  {Yevtushenko}},\ }\bibfield  {title} {\bibinfo {title} {Physics of
  arbitrarily doped kondo lattices: From a commensurate insulator to a heavy
  luttinger liquid and a protected helical metal},\ }\href
  {https://doi.org/10.1103/PhysRevB.100.165110} {\bibfield  {journal} {\bibinfo
   {journal} {Phys. Rev. B}\ }\textbf {\bibinfo {volume} {100}},\ \bibinfo
  {pages} {165110} (\bibinfo {year} {2019})}\BibitemShut {NoStop}%
\bibitem [{\citenamefont {Tsunetsugu}\ \emph {et~al.}(1997)\citenamefont
  {Tsunetsugu}, \citenamefont {Sigrist},\ and\ \citenamefont
  {Ueda}}]{Sigrist97}%
  \BibitemOpen
  \bibfield  {author} {\bibinfo {author} {\bibfnamefont {H.}~\bibnamefont
  {Tsunetsugu}}, \bibinfo {author} {\bibfnamefont {M.}~\bibnamefont
  {Sigrist}},\ and\ \bibinfo {author} {\bibfnamefont {K.}~\bibnamefont
  {Ueda}},\ }\bibfield  {title} {\bibinfo {title} {The ground-state phase
  diagram of the one-dimensional kondo lattice model},\ }\href
  {https://doi.org/10.1103/RevModPhys.69.809} {\bibfield  {journal} {\bibinfo
  {journal} {Rev. Mod. Phys.}\ }\textbf {\bibinfo {volume} {69}},\ \bibinfo
  {pages} {809} (\bibinfo {year} {1997})}\BibitemShut {NoStop}%
\bibitem [{\citenamefont {Basylko}\ \emph {et~al.}(2008)\citenamefont
  {Basylko}, \citenamefont {Lundow},\ and\ \citenamefont
  {Rosengren}}]{Basylko08}%
  \BibitemOpen
  \bibfield  {author} {\bibinfo {author} {\bibfnamefont {S.~A.}\ \bibnamefont
  {Basylko}}, \bibinfo {author} {\bibfnamefont {P.~H.}\ \bibnamefont
  {Lundow}},\ and\ \bibinfo {author} {\bibfnamefont {A.}~\bibnamefont
  {Rosengren}},\ }\bibfield  {title} {\bibinfo {title} {One-dimensional kondo
  lattice model studied through numerical diagonalization},\ }\href
  {https://doi.org/10.1103/PhysRevB.77.073103} {\bibfield  {journal} {\bibinfo
  {journal} {Phys. Rev. B}\ }\textbf {\bibinfo {volume} {77}},\ \bibinfo
  {pages} {073103} (\bibinfo {year} {2008})}\BibitemShut {NoStop}%
\bibitem [{\citenamefont {Lobos}\ \emph {et~al.}(2015)\citenamefont {Lobos},
  \citenamefont {Dobry},\ and\ \citenamefont {Galitski}}]{Lobos15}%
  \BibitemOpen
  \bibfield  {author} {\bibinfo {author} {\bibfnamefont {A.~M.}\ \bibnamefont
  {Lobos}}, \bibinfo {author} {\bibfnamefont {A.~O.}\ \bibnamefont {Dobry}},\
  and\ \bibinfo {author} {\bibfnamefont {V.}~\bibnamefont {Galitski}},\
  }\bibfield  {title} {\bibinfo {title} {Magnetic end states in a strongly
  interacting one-dimensional topological kondo insulator},\ }\href
  {https://doi.org/10.1103/PhysRevX.5.021017} {\bibfield  {journal} {\bibinfo
  {journal} {Phys. Rev. X}\ }\textbf {\bibinfo {volume} {5}},\ \bibinfo {pages}
  {021017} (\bibinfo {year} {2015})}\BibitemShut {NoStop}%
\bibitem [{\citenamefont {Zhong}\ \emph {et~al.}(2017)\citenamefont {Zhong},
  \citenamefont {Liu},\ and\ \citenamefont {Luo}}]{Zhong2017}%
  \BibitemOpen
  \bibfield  {author} {\bibinfo {author} {\bibfnamefont {Y.}~\bibnamefont
  {Zhong}}, \bibinfo {author} {\bibfnamefont {Y.}~\bibnamefont {Liu}},\ and\
  \bibinfo {author} {\bibfnamefont {H.-G.}\ \bibnamefont {Luo}},\ }\bibfield
  {title} {\bibinfo {title} {Topological phase in 1d topological kondo
  insulator: Z2 topological insulator, haldane-like phase and kondo
  breakdown},\ }\bibfield  {journal} {\bibinfo  {journal} {The European
  Physical Journal B}\ }\textbf {\bibinfo {volume} {90}},\ \href
  {https://doi.org/10.1140/epjb/e2017-80102-0} {10.1140/epjb/e2017-80102-0}
  (\bibinfo {year} {2017})\BibitemShut {NoStop}%
\bibitem [{\citenamefont {Zhong}\ \emph {et~al.}(2018)\citenamefont {Zhong},
  \citenamefont {Wang}, \citenamefont {Liu}, \citenamefont {Song},
  \citenamefont {Liu},\ and\ \citenamefont {Luo}}]{Zhong2018}%
  \BibitemOpen
  \bibfield  {author} {\bibinfo {author} {\bibfnamefont {Y.}~\bibnamefont
  {Zhong}}, \bibinfo {author} {\bibfnamefont {Q.}~\bibnamefont {Wang}},
  \bibinfo {author} {\bibfnamefont {Y.}~\bibnamefont {Liu}}, \bibinfo {author}
  {\bibfnamefont {H.-F.}\ \bibnamefont {Song}}, \bibinfo {author}
  {\bibfnamefont {K.}~\bibnamefont {Liu}},\ and\ \bibinfo {author}
  {\bibfnamefont {H.-G.}\ \bibnamefont {Luo}},\ }\bibfield  {title} {\bibinfo
  {title} {Finite temperature physics of 1d topological kondo insulator: Stable
  haldane phase, emergent energy scale and beyond},\ }\bibfield  {journal}
  {\bibinfo  {journal} {Frontiers of Physics}\ }\textbf {\bibinfo {volume}
  {14}},\ \href {https://doi.org/10.1007/s11467-018-0868-x}
  {10.1007/s11467-018-0868-x} (\bibinfo {year} {2018})\BibitemShut {NoStop}%
\bibitem [{\citenamefont {Khait}\ \emph {et~al.}(2018)\citenamefont {Khait},
  \citenamefont {Azaria}, \citenamefont {Hubig}, \citenamefont
  {Schollw{\"o}ck},\ and\ \citenamefont {Auerbach}}]{Khait5140}%
  \BibitemOpen
  \bibfield  {author} {\bibinfo {author} {\bibfnamefont {I.}~\bibnamefont
  {Khait}}, \bibinfo {author} {\bibfnamefont {P.}~\bibnamefont {Azaria}},
  \bibinfo {author} {\bibfnamefont {C.}~\bibnamefont {Hubig}}, \bibinfo
  {author} {\bibfnamefont {U.}~\bibnamefont {Schollw{\"o}ck}},\ and\ \bibinfo
  {author} {\bibfnamefont {A.}~\bibnamefont {Auerbach}},\ }\bibfield  {title}
  {\bibinfo {title} {Doped kondo chain, a heavy luttinger liquid},\ }\href
  {https://doi.org/10.1073/pnas.1719374115} {\bibfield  {journal} {\bibinfo
  {journal} {Proceedings of the National Academy of Sciences}\ }\textbf
  {\bibinfo {volume} {115}},\ \bibinfo {pages} {5140} (\bibinfo {year}
  {2018})},\ \Eprint
  {https://arxiv.org/abs/https://www.pnas.org/content/115/20/5140.full.pdf}
  {https://www.pnas.org/content/115/20/5140.full.pdf} \BibitemShut {NoStop}%
\bibitem [{\citenamefont {Trebst}\ \emph {et~al.}(2006)\citenamefont {Trebst},
  \citenamefont {Monien}, \citenamefont {Grzesik},\ and\ \citenamefont
  {Sigrist}}]{Trebst06}%
  \BibitemOpen
  \bibfield  {author} {\bibinfo {author} {\bibfnamefont {S.}~\bibnamefont
  {Trebst}}, \bibinfo {author} {\bibfnamefont {H.}~\bibnamefont {Monien}},
  \bibinfo {author} {\bibfnamefont {A.}~\bibnamefont {Grzesik}},\ and\ \bibinfo
  {author} {\bibfnamefont {M.}~\bibnamefont {Sigrist}},\ }\bibfield  {title}
  {\bibinfo {title} {Quasiparticle dynamics in the kondo lattice model at half
  filling},\ }\href {https://doi.org/10.1103/PhysRevB.73.165101} {\bibfield
  {journal} {\bibinfo  {journal} {Phys. Rev. B}\ }\textbf {\bibinfo {volume}
  {73}},\ \bibinfo {pages} {165101} (\bibinfo {year} {2006})}\BibitemShut
  {NoStop}%
\bibitem [{\citenamefont {Smerat}\ \emph {et~al.}(2009)\citenamefont {Smerat},
  \citenamefont {Schollw\"ock}, \citenamefont {McCulloch},\ and\ \citenamefont
  {Schoeller}}]{Smerat09}%
  \BibitemOpen
  \bibfield  {author} {\bibinfo {author} {\bibfnamefont {S.}~\bibnamefont
  {Smerat}}, \bibinfo {author} {\bibfnamefont {U.}~\bibnamefont
  {Schollw\"ock}}, \bibinfo {author} {\bibfnamefont {I.~P.}\ \bibnamefont
  {McCulloch}},\ and\ \bibinfo {author} {\bibfnamefont {H.}~\bibnamefont
  {Schoeller}},\ }\bibfield  {title} {\bibinfo {title} {Quasiparticles in the
  kondo lattice model at partial fillings of the conduction band using the
  density matrix renormalization group},\ }\href
  {https://doi.org/10.1103/PhysRevB.79.235107} {\bibfield  {journal} {\bibinfo
  {journal} {Phys. Rev. B}\ }\textbf {\bibinfo {volume} {79}},\ \bibinfo
  {pages} {235107} (\bibinfo {year} {2009})}\BibitemShut {NoStop}%
\bibitem [{\citenamefont {Appelbaum}(1966)}]{Appelbaum1966}%
  \BibitemOpen
  \bibfield  {author} {\bibinfo {author} {\bibfnamefont {J.}~\bibnamefont
  {Appelbaum}},\ }\bibfield  {title} {\bibinfo {title} {"$s\ensuremath{-}d$"
  exchange model of zero-bias tunneling anomalies},\ }\href
  {https://doi.org/10.1103/PhysRevLett.17.91} {\bibfield  {journal} {\bibinfo
  {journal} {Phys. Rev. Lett.}\ }\textbf {\bibinfo {volume} {17}},\ \bibinfo
  {pages} {91} (\bibinfo {year} {1966})}\BibitemShut {NoStop}%
\bibitem [{\citenamefont {Anderson}(1966)}]{Anderson1966}%
  \BibitemOpen
  \bibfield  {author} {\bibinfo {author} {\bibfnamefont {P.~W.}\ \bibnamefont
  {Anderson}},\ }\bibfield  {title} {\bibinfo {title} {Localized magnetic
  states and fermi-surface anomalies in tunneling},\ }\href
  {https://doi.org/10.1103/PhysRevLett.17.95} {\bibfield  {journal} {\bibinfo
  {journal} {Phys. Rev. Lett.}\ }\textbf {\bibinfo {volume} {17}},\ \bibinfo
  {pages} {95} (\bibinfo {year} {1966})}\BibitemShut {NoStop}%
\bibitem [{\citenamefont {Pustilnik}\ and\ \citenamefont
  {Glazman}(2001)}]{Pustilnik01}%
  \BibitemOpen
  \bibfield  {author} {\bibinfo {author} {\bibfnamefont {M.}~\bibnamefont
  {Pustilnik}}\ and\ \bibinfo {author} {\bibfnamefont {L.~I.}\ \bibnamefont
  {Glazman}},\ }\bibfield  {title} {\bibinfo {title} {Kondo effect in real
  quantum dots},\ }\href {https://doi.org/10.1103/PhysRevLett.87.216601}
  {\bibfield  {journal} {\bibinfo  {journal} {Phys. Rev. Lett.}\ }\textbf
  {\bibinfo {volume} {87}},\ \bibinfo {pages} {216601} (\bibinfo {year}
  {2001})}\BibitemShut {NoStop}%
\bibitem [{\citenamefont {Maltseva}\ \emph {et~al.}(2009)\citenamefont
  {Maltseva}, \citenamefont {Dzero}, \citenamefont {Dzero}, \citenamefont
  {Coleman},\ and\ \citenamefont {Coleman}}]{Maltseva2009}%
  \BibitemOpen
  \bibfield  {author} {\bibinfo {author} {\bibfnamefont {M.}~\bibnamefont
  {Maltseva}}, \bibinfo {author} {\bibfnamefont {M.}~\bibnamefont {Dzero}},
  \bibinfo {author} {\bibfnamefont {M.}~\bibnamefont {Dzero}}, \bibinfo
  {author} {\bibfnamefont {P.}~\bibnamefont {Coleman}},\ and\ \bibinfo {author}
  {\bibfnamefont {P.}~\bibnamefont {Coleman}},\ }\bibfield  {title} {\bibinfo
  {title} {Electron cotunneling into a kondo lattice},\ }\href
  {https://doi.org/10.1103/physrevlett.103.206402} {\bibfield  {journal}
  {\bibinfo  {journal} {Physical Review Letters}\ }\textbf {\bibinfo {volume}
  {103}},\ \bibinfo {pages} {206402} (\bibinfo {year} {2009})}\BibitemShut
  {NoStop}%
\bibitem [{\citenamefont {Schollw{\"o}ck}(2011)}]{Schollwoeck11}%
  \BibitemOpen
  \bibfield  {author} {\bibinfo {author} {\bibfnamefont {U.}~\bibnamefont
  {Schollw{\"o}ck}},\ }\bibfield  {title} {\bibinfo {title} {The density-matrix
  renormalization group in the age of matrix product states},\ }\href
  {https://doi.org/https://doi.org/10.1016/j.aop.2010.09.012} {\bibfield
  {journal} {\bibinfo  {journal} {Annals of Physics}\ }\textbf {\bibinfo
  {volume} {326}},\ \bibinfo {pages} {96} (\bibinfo {year} {2011})}\BibitemShut
  {NoStop}%
\bibitem [{\citenamefont {Vidal}(2004)}]{Vidal04}%
  \BibitemOpen
  \bibfield  {author} {\bibinfo {author} {\bibfnamefont {G.}~\bibnamefont
  {Vidal}},\ }\bibfield  {title} {\bibinfo {title} {Efficient simulation of
  one-dimensional quantum many-body systems},\ }\href
  {https://doi.org/10.1103/PhysRevLett.93.040502} {\bibfield  {journal}
  {\bibinfo  {journal} {Phys. Rev. Lett.}\ }\textbf {\bibinfo {volume} {93}},\
  \bibinfo {pages} {040502} (\bibinfo {year} {2004})}\BibitemShut {NoStop}%
\bibitem [{\citenamefont {Daley}\ \emph {et~al.}(2004)\citenamefont {Daley},
  \citenamefont {Kollath}, \citenamefont {Schollw{\"o}ck},\ and\ \citenamefont
  {Vidal}}]{Daley04}%
  \BibitemOpen
  \bibfield  {author} {\bibinfo {author} {\bibfnamefont {A.~J.}\ \bibnamefont
  {Daley}}, \bibinfo {author} {\bibfnamefont {C.}~\bibnamefont {Kollath}},
  \bibinfo {author} {\bibfnamefont {U.}~\bibnamefont {Schollw{\"o}ck}},\ and\
  \bibinfo {author} {\bibfnamefont {G.}~\bibnamefont {Vidal}},\ }\bibfield
  {title} {\bibinfo {title} {Time-dependent density-matrix
  renormalization-group using adaptive effective hilbert spaces},\ }\href@noop
  {} {\bibfield  {journal} {\bibinfo  {journal} {Journal of Statistical
  Mechanics: Theory and Experiment}\ }\textbf {\bibinfo {volume} {2004}},\
  \bibinfo {pages} {P04005} (\bibinfo {year} {2004})}\BibitemShut {NoStop}%
\bibitem [{\citenamefont {White}\ and\ \citenamefont
  {Feiguin}(2004)}]{White04}%
  \BibitemOpen
  \bibfield  {author} {\bibinfo {author} {\bibfnamefont {S.~R.}\ \bibnamefont
  {White}}\ and\ \bibinfo {author} {\bibfnamefont {A.~E.}\ \bibnamefont
  {Feiguin}},\ }\bibfield  {title} {\bibinfo {title} {Real-time evolution using
  the density matrix renormalization group},\ }\href
  {https://doi.org/10.1103/PhysRevLett.93.076401} {\bibfield  {journal}
  {\bibinfo  {journal} {Phys. Rev. Lett.}\ }\textbf {\bibinfo {volume} {93}},\
  \bibinfo {pages} {076401} (\bibinfo {year} {2004})}\BibitemShut {NoStop}%
\bibitem [{\citenamefont {Fishman}\ \emph {et~al.}(2022)\citenamefont
  {Fishman}, \citenamefont {White},\ and\ \citenamefont
  {Stoudenmire}}]{itensor}%
  \BibitemOpen
  \bibfield  {author} {\bibinfo {author} {\bibfnamefont {M.}~\bibnamefont
  {Fishman}}, \bibinfo {author} {\bibfnamefont {S.~R.}\ \bibnamefont {White}},\
  and\ \bibinfo {author} {\bibfnamefont {E.~M.}\ \bibnamefont {Stoudenmire}},\
  }\bibfield  {title} {\bibinfo {title} {{The ITensor Software Library for
  Tensor Network Calculations}},\ }\href
  {https://doi.org/10.21468/SciPostPhysCodeb.4} {\bibfield  {journal} {\bibinfo
   {journal} {SciPost Phys. Codebases}\ ,\ \bibinfo {pages} {4}} (\bibinfo
  {year} {2022})}\BibitemShut {NoStop}%
\bibitem [{\citenamefont {Oshikawa}(2000)}]{Oshikawa2000}%
  \BibitemOpen
  \bibfield  {author} {\bibinfo {author} {\bibfnamefont {M.}~\bibnamefont
  {Oshikawa}},\ }\bibfield  {title} {\bibinfo {title} {Topological approach to
  luttinger's theorem and the fermi surface of a kondo lattice},\ }\href
  {https://doi.org/10.1103/PhysRevLett.84.3370} {\bibfield  {journal} {\bibinfo
   {journal} {Phys. Rev. Lett.}\ }\textbf {\bibinfo {volume} {84}},\ \bibinfo
  {pages} {3370} (\bibinfo {year} {2000})}\BibitemShut {NoStop}%
\bibitem [{\citenamefont {Martin}(1982)}]{Martin82}%
  \BibitemOpen
  \bibfield  {author} {\bibinfo {author} {\bibfnamefont {R.~M.}\ \bibnamefont
  {Martin}},\ }\bibfield  {title} {\bibinfo {title} {Fermi-surface sum rule and
  its consequences for periodic {K}ondo and mixed-valence systems},\ }\href
  {https://link.aps.org/doi/10.1103/PhysRevLett.48.362} {\bibfield  {journal}
  {\bibinfo  {journal} {Phys. Rev. Lett.}\ }\textbf {\bibinfo {volume} {48}},\
  \bibinfo {pages} {362} (\bibinfo {year} {1982})}\BibitemShut {NoStop}%
\bibitem [{\citenamefont {Fradkin}\ and\ \citenamefont
  {Shenker}(1979)}]{ShenkerFradkin}%
  \BibitemOpen
  \bibfield  {author} {\bibinfo {author} {\bibfnamefont {E.}~\bibnamefont
  {Fradkin}}\ and\ \bibinfo {author} {\bibfnamefont {S.~H.}\ \bibnamefont
  {Shenker}},\ }\bibfield  {title} {\bibinfo {title} {Phase diagrams of lattice
  gauge theories with higgs fields},\ }\href
  {https://doi.org/10.1103/PhysRevD.19.3682} {\bibfield  {journal} {\bibinfo
  {journal} {Phys. Rev. D}\ }\textbf {\bibinfo {volume} {19}},\ \bibinfo
  {pages} {3682} (\bibinfo {year} {1979})}\BibitemShut {NoStop}%
\bibitem [{\citenamefont {Hazra}\ and\ \citenamefont
  {Coleman}(2021)}]{Hazra21}%
  \BibitemOpen
  \bibfield  {author} {\bibinfo {author} {\bibfnamefont {T.}~\bibnamefont
  {Hazra}}\ and\ \bibinfo {author} {\bibfnamefont {P.}~\bibnamefont
  {Coleman}},\ }\bibfield  {title} {\bibinfo {title} {Luttinger sum rules and
  spin fractionalization in the su($n$) kondo lattice},\ }\href
  {https://doi.org/10.1103/PhysRevResearch.3.033284} {\bibfield  {journal}
  {\bibinfo  {journal} {Phys. Rev. Res.}\ }\textbf {\bibinfo {volume} {3}},\
  \bibinfo {pages} {033284} (\bibinfo {year} {2021})}\BibitemShut {NoStop}%
\bibitem [{\citenamefont {Ge}\ and\ \citenamefont {Komijani}(2022)}]{Ge2022}%
  \BibitemOpen
  \bibfield  {author} {\bibinfo {author} {\bibfnamefont {Y.}~\bibnamefont
  {Ge}}\ and\ \bibinfo {author} {\bibfnamefont {Y.}~\bibnamefont {Komijani}},\
  }\bibfield  {title} {\bibinfo {title} {Emergent spinon dispersion and
  symmetry breaking in two-channel kondo lattices},\ }\href
  {https://doi.org/10.1103/PhysRevLett.129.077202} {\bibfield  {journal}
  {\bibinfo  {journal} {Phys. Rev. Lett.}\ }\textbf {\bibinfo {volume} {129}},\
  \bibinfo {pages} {077202} (\bibinfo {year} {2022})}\BibitemShut {NoStop}%
\bibitem [{\citenamefont {Wugalter}\ \emph {et~al.}(2020)\citenamefont
  {Wugalter}, \citenamefont {Komijani},\ and\ \citenamefont
  {Coleman}}]{Wugalter2020}%
  \BibitemOpen
  \bibfield  {author} {\bibinfo {author} {\bibfnamefont {A.}~\bibnamefont
  {Wugalter}}, \bibinfo {author} {\bibfnamefont {Y.}~\bibnamefont {Komijani}},\
  and\ \bibinfo {author} {\bibfnamefont {P.}~\bibnamefont {Coleman}},\
  }\bibfield  {title} {\bibinfo {title} {Large-$n$ approach to the two-channel
  kondo lattice},\ }\href {https://doi.org/10.1103/PhysRevB.101.075133}
  {\bibfield  {journal} {\bibinfo  {journal} {Phys. Rev. B}\ }\textbf {\bibinfo
  {volume} {101}},\ \bibinfo {pages} {075133} (\bibinfo {year}
  {2020})}\BibitemShut {NoStop}%
\end{thebibliography}%

\end{document}